\documentclass[10pt]{amsart} 
\usepackage{amsmath, amsfonts, amsthm, amssymb, verbatim,dsfont}  
\usepackage{graphicx}
\usepackage[mathcal]{euscript} 
\usepackage{amstext} 
\usepackage{amssymb,mathrsfs} 
\theoremstyle{plain} 
\theoremstyle{definition} 
\theoremstyle{plain}
          
\numberwithin{equation}{section} 
\newcommand \auth {\textsc} 
\newcommand \jou {\textit}
\newcommand \be           {\begin{equation}}
\newcommand \ee            {\end{equation}}

\newcommand \RR           {\mathbb{R}} 

\newcommand \ZZ           {\mathbb{Z}}

\newcommand \del           \partial
\newcommand \eps        	\epsilon

\newcommand \la 		\langle
\newcommand \ra 		\rangle
\begin{document}

\title[Computing Gowdy spacetimes in future and past directions]
{\small Computing Gowdy spacetimes via spectral evolution 
in future and past directions}

\author
%            correct spelling :        "LeFloch"  or  "LeFLOCH" 
% 
[\textsc{P. Amorim, C. Bernardi,} \and \textsc{P.G. L{\tiny e}Floch}]
{\textsc{Paulo Amorim, Christine Bernardi,} 
\\
\and 
\\
\textsc{Philippe G. L{\tiny e}Floch}}
\address
{
%\textsc{P. Amorim, C. Bernardi,} \and \textsc{P.G. LeFloch}\\
Laboratoire Jacques-Louis Lions \& Centre National de la Recherche Scientifique (CNRS)
\\
Universit\'e Pierre et Marie Curie (Paris 6), 4 Place Jussieu, 75252 Paris, France.
\newline 
E-mail: \textit{Amorim@ann.jussieu.fr, Bernardi@ann.jussieu.fr, pgLeFloch@gmail.com}}
 
\date{\today}

\subjclass[2000]   {Primary : 65M70, 65T40.     Secondary :  83C05, 35L05.
{\it Key words and phrases.}
 Einstein equations, Gowdy symmetry, pseudo-spectral method, future or past evolution.
\newline
To appear in: {\it Classical and Quantum Gravity.}
}

\begin{abstract} We consider a system of nonlinear wave equations with constraints
that arises from the Einstein equations of general relativity and 
describes the geometry of the so-called Gowdy symmetric 
spacetimes on $T^3$. We introduce two numerical methods, which are based on pseudo-spectral approximation.    
The first approach relies on marching in the future time-like direction and toward the coordinate 
singularity $t=0$. The second approach is designed from asymptotic formulas that are available near this singularity; 
it evolves the solutions in the past timelike direction from ``final'' data given at $t=0$.
This backward method relies a novel nonlinear transformation, 
which allows us to reduce the nonlinear source terms to simple quadratic products of the unknown variables. 
Numerical experiments are presented in various regimes, including cases where ``spiky'' structures are observed 
as the singularity is approached.  The proposed backward strategy leads to a robust numerical 
method which allows us to accurately simulate the long-time 
behavior of a large class of Gowdy spacetimes.
\end{abstract}
\maketitle

%===========================================================================
\section{Introduction}
\label{GIN}

We consider spacetimes $(M,g)$ 
foliated by spatial slices that have the topology of the $3$-torus $T^3$ 
and on which a two-parameter group of isometries acts. As shown by Gowdy \cite{Gowdy}, a 
large class of such spacetimes 
can be described in suitably chosen coordinates $(\tau, \theta,\alpha,\beta) \in \RR \times T^3$ 
in which the metric $g$ reads 
$$
g = e^{(\tau - \lambda)/2} \, ( -e^{-2\tau} d\tau^2 + d \theta^2)
	+ e^{-\tau} \, \big( e^P \, d \alpha^2 + 2 e^P Q \, d\alpha d\beta +
	(e^P Q^2 + e^{-P}) \, d \beta^2 \big).
$$
The unknown metric coefficients $P, Q, \lambda$ depend upon the variables $\tau, \theta$, only, 
and are periodic in the spatial variable $\theta$.  

In absence of matter fields,
the Einstein equations for Gowdy spacetimes are equivalent to two 
second-order, nonlinear wave equations for the time-evolution of $P,Q$: 
\be
\aligned
& P_{\tau \tau} - e^{-2 \tau} P_{\theta\theta} = e^{2P} \, (Q^2_\tau - e^{-2\tau} Q^2_\theta), 
\\
& Q_{\tau\tau} - e^{-2 \tau} Q_{\theta\theta}  = - 2(P_\tau Q_\tau - e^{-2\tau} P_\theta Q_\theta), 
\endaligned
\label{GS.Einstein1} 
\ee
together with first-order differential equations for the coefficient $\lambda$:  
\be
\aligned
& \lambda_\tau = P^2_\tau + e^{-2\tau} P^2_\theta + e^{2P} \, (Q^2_\tau + e^{-2\tau} Q^2_\theta), 
\\
& \lambda_\theta = 2(P_\tau P_\theta + e^{2P} Q_\tau Q_\theta). 
\endaligned 
\label{GS.Einstein2} 
\ee 
The functions $P,Q$ can be determined first by solving \eqref{GS.Einstein1}, 
and, then, the function $\lambda$ 
can be obtained by integrating \eqref{GS.Einstein2}. Since, by assumption, $\lambda$ is a
periodic function, we have 
$\int_0^{2\pi} \lambda_\theta \, d \theta = 0$ and, therefore, the functions $P,Q$ 
must satisfy the constraint
\be
\int_0^{2\pi} (P_\tau P_\theta + e^{2P} Q_\tau Q_\theta) \, d\theta = 0. 
\label{GS.Einstein3} 
\ee 
In fact, it is easily checked that if \eqref{GS.Einstein3} holds at some positive time $\tau_0$, 
then it holds for all positive times. 

The properties of Gowdy spacetimes have been extensively studied in recent years, both 
theoretically and numerically. 
We refer the reader to the works by Garfinkle \cite{Garfinkle}, Isenberg \cite{Isenberg}, 
Isenberg and Moncrief \cite{IsenbergMoncrief0,IsenbergMoncrief}
and Rendall \cite{Rendall0,Rendall}, and Rendall and Weaver \cite{RendallWeaver}, 
and the references cited therein. 
Further properties of Gowdy spacetimes have been established by Chae and 
Chrusciel \cite{CC} and Chrusciel and Lake \cite{CL}. 
The structure of Gowdy spacetimes was completely solved in a series of 
works by Ringstr\"om \cite{Ringstrom1}--\cite{Ringstrom6}. 
In particular, \cite{Ringstrom3} gives a mathematical analysis of the asymptotics in
the expanding direction for $T^3$ Gowdy metrics, and \cite{Ringstrom6} 
(building on the other papers) provides a complete mathematical description
of the asymptotic behaviour for generic solutions as one approaches the
singularity. 
Issues related to curvature blow-up are also well understood 
and generically $t=0$ corresponds to a curvature singularity. 

Our purpose in the present paper is restricted to presenting a new strategy for {\sl computing} Gowdy spacetimes, 
and demontrating its interest for analyzing the behavior of spacetimes near the coordinate singularity. 
This strategy takes its roots in an approach developed by Rendall \cite{Rendall} jointly 
with Kichenessamy \cite{KR} which led to {\sl existence} results for the above equations. 
Our proposed numerical approach will be based on spectral approximations whose applicability to
problems arising in general relativity is now well-recognized; see the works by Ansorg \cite{Ansorg}, 
Bonazzola, Gourgoulhon, Grandcl\'ement, and Novak \cite{BGGN}, and Lehner, Reula, and Tiglio \cite{LRT}
and the references cited therein.

One of the main issues of interest is to elucidate the qualitative behavior 
of solutions to the Einstein equations \eqref{GS.Einstein1} and, especially, 
to understand whether these solutions blow-up in the past timelike direction
 $\tau \to + \infty$. 
By contrast, in the future direction $\tau \to -\infty$ the solution remains globally regular
and the dynamics is better understood.   
We are here mainly interested in the region $\tau \in [\tau_0,\infty)$ and, without loss of generality, 
we take $\tau_0 =0$. By 
 introducing the rescaled time 
$$
t = e^{-\tau} \in (0,1],
$$
the system \eqref{GS.Einstein1} becomes 
\be
\aligned
& P_{tt} + {P_t \over t} - P_{\theta\theta} = e^{2P} ( Q^2_t - Q^2_\theta),
\\
& Q_{tt} + {Q_t \over t} - Q_{\theta\theta} = -2(P_t Q_t - P_\theta Q_\theta). 
\endaligned
\label{GS.Einstein4}
\ee
For these equations, 
we will consider the initial-value problem when  
data are prescribed on either the hypersurface $\tau=0$ 
(forward problem),
or the singularity $t=0$ (backward problem).  

Let us end this introduction with a brief discussion of formal expansions for 
the solutions of \eqref{GS.Einstein1} and some heuristics arguments. 

According to \cite{Garfinkle,IsenbergMoncrief,Rendall}, as one approaches the coordinate singularity, 
the spatial derivative terms in \eqref{GS.Einstein4} becomes negligible and $P,Q,$ 
approach solutions of the \emph{ordinary} differential equations 
\be
\aligned
& P_{tt} + {P_t \over t} = e^{2P} Q^2_t, 
\\
& Q_{tt} + {Q_t \over t} = - 2 \, P_t \, Q_t.
\endaligned
\label{GS.ODE}
\ee
These equations are referred to as the \emph{velocity term dominated} (VTD) equations.  
Interestingly, the VTD equations \eqref{GS.ODE} admit a large class of 
explicit solutions \cite{Garfinkle}
and the asymptotic analysis of these solutions yields the following behavior (as $t\to 0$) 
\be
\aligned
& P = - v \, \log t + \varphi + o(t),
\\
& Q = q + t^{2 v} \, ( \psi + o(1)). 
\endaligned
\label{GS.expa}
\ee
Here, $v, \varphi, q$ and $\psi$ are prescribed coefficients depending on the parameter $\theta$,
and the function $v$ is referred to as the \emph{asymptotic velocity}.

On the other hand, solutions to the full nonlinear equations \eqref{GS.Einstein4} 
have been constructed \cite{Rendall} which have an asymptotic behavior 
described by \eqref{GS.expa} and are called  \emph{asymptotically velocity term dominated} (AVTD) solutions. 

It is important to observe the analysis leading to solutions of \eqref{GS.Einstein4} 
satisfying \eqref{GS.expa} put drastic constrains for 
the range of the asymptotic velocity $v$. In fact, the function $v$ must range within
the interval $(0,1)$, except possibly at some exceptional points 
where the coefficients $q_\theta$ or $\psi$ vanish. At these points, the solutions may develop
\emph{spikes}; we will discuss the properties of these spikes later in the present paper. 
The analysis is confirmed by numerous numerical experiments \cite{Garfinkle}.

An outline of this paper follows.  
In Section~\ref{FM}, we introduce a pseudo-spectral approximation scheme for 
solving the forward initial-value problem \eqref{GS.Einstein1}. 
Then, in Section~\ref{BM}, we turn our attention to the backward initial-value problem for \eqref{GS.Einstein4},
and we introduce an adapted pseudo-spectral approximation for this problem. 
Finally, in Section~\ref{NE} we present extensive numerical experiments; in particular, 
by running a backward evolution followed by a forward evolution, we can demonstrate the accuracy of the proposed
method and investigate the behavior of Gowdy spacetimes near the coordinate singularity.  
 
%=============================================================================================================

\section{Forward method based on pseudo-spectral approximation}
\label{FM}

In spectral approximation methods, one expands the unknown function in an appropriate basis 
(polynomials, trigonometric polynomials, etc.) and one numerically solves the coupled system of
equations satisfied by the components of the solution relative to the chosen basis. Here, we 
choose Fourier 
expansions, which are natural since we are dealing with periodic functions.

Before we discretize the system \eqref{GS.Einstein1} under consideration, we propose here 
to recast it in a first-order form, 
by using a nonlinear transformation of the dependent variables: 
the proposed transformation allows us to reduce the nonlinear source terms 
to simple quadratic products of the unknown variables. Defining
$$
\aligned
& V := P_\tau, \qquad && W := e^{-\tau} P_\theta,
& X := e^{P-\tau} Q_\theta, \qquad && Y:= e^P Q_\tau, 
\endaligned
$$
the Einstein equations \eqref{GS.Einstein1} in these new variables read 
\be
\label{SN.5}
\aligned
& V_\tau - e^{-\tau} W_\theta = Y^2 - X^2,
\\
& W_\tau - e^{-\tau} V_\theta = - W,
\\
& X_\tau - e^{-\tau} Y_\theta = - (1-V) X  - W Y,
\\
& Y_\tau - e^{-\tau} X_\theta = -V Y + W X.
\endaligned
\ee
Observe that the right-hand sides of the above equations contain both linear and 
quadratic terms. 

We rely on the proposed formulation \eqref{SN.5} and introduce Fourier series expansions
of the $2\pi$-periodic, unknown functions $V,W,X,Y$, that is,  we set 
\be
\label{SN.1}
\aligned
V(\tau,\theta) = \sum_{k\in\ZZ} V^k(\tau) e^{ik\theta},
\endaligned
\ee
and analogously for the other unknowns. The Fourier coefficient $V^k$
is given by 
\be
\label{SN.1.1}
\aligned
V^k(\tau) =\frac{1}{2\pi} \int_{-\pi}^\pi V(\tau,\theta) e^{-ik\theta} d\theta
\endaligned
\ee
while the $\theta$-derivative of $V$ is obtained by differentiating each term of the above expansion: 
$$
V_{\theta}(\tau,\theta) = \sum_{k\in\ZZ} V^k(\tau)ik\, e^{ik\theta}.
$$
Then, the system \eqref{GS.Einstein1} can be written as an infinite system of 
ordinary differential equations (ODE's) for the Fourier coefficients $V^k,W^k,X^k,Y^k$, that is, 
\be
\label{SN.60}
\aligned
& V^k_\tau - e^{-\tau}ik \, W^k = (Y^2 - X^2)^k =: S^k_1,
\\
& W^k_\tau - e^{-\tau}ik \, V^k = - W^k =: S^k_2,
\\
& X^k_\tau - e^{-\tau}ik \, Y^k = - X^k + (V X)^k  - (W Y)^k =: S^k_3,
\\
& Y^k_\tau - e^{-\tau}ik \, X^k = - (V Y)^k + (W X)^k =: S^k_4. 
\endaligned
\ee 
The nonlinear source-terms $S_i^k$ above are (at most) {\sl quadratic} in the unknowns and, therefore, 
they are easily be computed via the relation 
$$
\aligned
(AB)^k := \sum_{l\in\ZZ}A^{k-l}B^l. 
\endaligned
$$
At this stage, we obtain the (exact!) expressions 
\be
\label{SN.70}
\aligned
& S^k_1 = \sum_{l\in\ZZ}( Y^l Y^{k-l} - X^l X^{k-l}), 
\\
& S^k_2 =  - W^k,  
\\
&S^k_3 =  - X^k + \sum_{l\in\ZZ}(V^l  X^{k-l}  - W^l Y^{k-l}), 
\\
&S^k_4 = \sum_{l\in\ZZ}( - V^l  Y^{k-l} + W^l  X^{k-l}). 
\endaligned
\ee

In order to arrive at a practical numerical method, two further approximations are necessary.
On one hand, only a finite number of frequencies $k$ can be dealt with in practice and, therefore, 
we should truncate the sum in \eqref{SN.1} and 
by fixing a (sufficiently large) integer $K$ we obtain the following approximation of the function $V$   
\be
\label{167} 
V^{[K]}(\tau,\theta) := \sum_{|k|\le K} V^k(\tau) e^{ik\theta}, 
\ee
and we proceed similarly for the other unknowns $W,X,Y$. This approximation leads to
what is referred to as the {\sl spectral method.} 

On the other hand, we observe that the above approach relies on exact Fourier coefficients of the unknown functions, and 
we introduce an approximation of the integrals in \eqref{SN.1.1} 
based, specifically, on a quadrature formula involving the points $\theta_j = \frac{2\pi j}{2K+1}$. 
Precisely, {\sl instead of}
 \eqref{SN.1}, we actually define the (approximate) Fourier coefficients as the following averages: 
\be
\label{SN.1.2}
\aligned
&V^k_*(\tau) := \frac{1}{2K+1} \sum_{|j|\le K} V(\tau, \theta_j) e^{-ik\theta_j},
\endaligned
\ee
and, instead of \eqref{167}, we set
\be
\label{168} 
V_*^{[K]}(\tau,\theta) := \sum_{|k|\le K} V_*^k(\tau) e^{ik\theta}.
\ee
We proceed similarly for the other unknowns of the system of equations under consideration 
This approach is referred to as the \textsl{pseudo-spectral method}. 

Note in passing that the above method may be viewed as a {\sl collocation method.} 
Namely, approximating the Fourier coefficients by \eqref{SN.1.2} 
is equivalent to approximating the function $V$ itself by a trigonometric polynomial
which coincides with $V$ on the collocation points $\theta_j$. 
For further details and for the corresponding theory of convergence, 
we refer the reader to the textbook by Bernardi and Maday \cite{BM}.

Now, let us fix an integer $K$. Given suitable initial data for our problem, that is, 
the $2\pi$-periodic functions $\theta \to (V,W,X,Y)(0,\theta)$ satisfying the Einstein constraints, we can introduce
(for all $k=-K,\dots, K$)  
$$
\aligned 
&(V,W,X,Y)_0^k := (V,W,X,Y)^k_*(0),
\endaligned
$$
as defined in \eqref{SN.1.2} from the given initial data.
Then, we compute for instance
$$
\aligned
&S_{1,0}^k : = \sum_{|l|, |k-l| \le K}( Y_0^l Y_0^{k-l} - X_0^l X_0^{k-l}),
\endaligned
$$
and proceed similarly for the other source-terms. The range in the above summation is determined so that 
only frequencies that are below $K$ (in modulus) are taken into account. 
 
Next, using the above expressions of the source-terms, we let successively $n=1,2, \ldots$ and 
we evolve the Fourier coefficients by numerically solving \eqref{SN.60}
with a standard ODE solver. 
For the numerical results presented in Section~\ref{NE} below, 
we have implemented both a fourth-order Runge--Kutta method and the explicit Euler method and finally adopted 
the latter since the results did not differ significantly. 
The time-discretization is defined as follows. We denote by $\Delta\tau$ the time-increment
and set $\tau_n := n \Delta\tau$. Writing $A$ instead of $V,W,X,Y$, and $B$ instead of $W,V,Y,X$, respectively,
we can write the Euler scheme in the form ($i=1,2,3,4$) 
$$
A^k_{n+1} = A^k_n + \Delta\tau(e^{-\tau_n} ik B^k_n + S^k_{i,n}). 
$$

%===============================================================================================================

\section{Backward method based on pseudo-spectral approximation}
\label{BM}  

\subsection{Motivation for the backward strategy}

We now describe the new approach proposed in this paper.  
In the previous section, we described a method which allows us to calculate the metric coefficients 
$P,Q$ numerically, by evolving from a {\sl non-singular} coordinate time \emph{towards} the 
singularity. We refer to this approach as the \emph{forward method}.

We now introduce an alternative strategy and consider the problem \eqref{GS.Einstein1} 
as a \emph{backward problem} with data prescribed \emph{on} the singularity. 
This allows us to simulate the equations \eqref{GS.Einstein1} (now written in the form \eqref{GS.Einstein4}) 
in the reverse time-direction. 

The behavior of solutions may then be investigated by evolving either towards the singularity 
(using the forward method) or away from it (using the backward method).
Moreover,  
having two strategies allows us to evaluate the consistency of both methods
by, first, evolving away from the coordinate singularity and, then, using the resulting solution as an 
initial data for the forward method. 
As it will turn out, this backward-forward calculation provide numerical values on the initial slice
that are close to the initial data originally prescribed on the slice, and 
our general proposal therefore provides a reliable tool to investigate the behavior of Gowdy spacetimes, 
especially near the singularity.

It is important to stress that performing this backward-forward evolution {\sl does not} amount 
to simply computing the solution from $t=0$ to $t=1$ and, then, ``undoing'' the computation 
in the opposite direction. Indeed, the two systems of equations involved in the numerical 
simulation are highly nonlinear and, furthermore,  
they evolve the solution in opposite time directions. 

Another validation of the proposed backward strategy is given by the fact that, as we will see,  
the numerical results of our forward simulations are virtually indistinguishable from the ones 
found in the literature. See Section~\ref{NE} below for further details.

In order to move away from the singularity, we start by considering the Einstein equations 
in the form \eqref{GS.Einstein4}. However, if we try to convert this 
into a first-order system in the same way as in the previous section, we obtain 
\[
\aligned
& V_t -  W_\theta = \frac{1}{t} (X^2 - Y^2 ), \qquad && W_t -  V_\theta = \frac{1}{t} W,
\\
& X_t -  Y_\theta = \frac{1}{t} \big( (1-V) X  +W Y \big), \qquad && Y_t -  X_\theta = \frac{1}{t} (V Y - W X).
\endaligned
\]
This system is not amenable to stable numerical simulations as it stands, due to the singularities $1/t$ 
arising in the right-hand sides. Indeed, if we were applying the Euler method and evolve the solutions
in time, the scheme {\sl would not depend} on the time-increment. Moreover, as can be seen from the asymptotic 
expansions, three of these four quantities vanish (in general) on the 
coordinate singularity. Therefore, it is not possible to use the above formulation to numerically evolve from the
coordinate singularity.

%---------------------------------------------------------------------------------------------------

\subsection{Rescaled initial-value problem}

Hence a new approach, presented now, is needed to successfully evolve from the singularity. Recall that the behavior of the metric coefficients is determined by the main coefficients $v,\varphi,\psi,q$ of the asymptotic expansions of $P,Q$. 
This motivate us to introduce the rescaled unknowns
\be
\label{BCP.8}
\aligned
& V = -t P_t, \quad        &&\Phi = P + v\log t,
\\
& \Psi = t^{1-2v} Q_t,    \quad && \chi = Q_\theta,
\endaligned
\ee
which, according to the asymptotic expansions \eqref{GS.expa}, are expected to have a limit on the 
singularity and to converge toward the given data $v,\varphi,\psi,q$ as $t \to 0$: 
$$
\aligned
& V(t,\theta) \to v(\theta), \quad  && \Phi(t,\theta) \to \varphi(\theta),
\\
& \Psi(t,\theta) \to 2v(\theta) \psi(\theta), \quad  && \chi(t,\theta) \to q'(\theta).
\endaligned
$$

Expressing the Einstein equations as a system of evolution equations in terms of the proposed unknowns 
yields the system
\be
\label{BCP.10}
\aligned
& V_t = -t \Phi_{\theta\theta} + t v''\log t 
        -e^{2\Phi}\big( t^{2v -1} \Psi^2 -  t^{1-2v}\chi^2 \big),
\\
& \Phi_t = \frac{v - V}{t},
\\
& \Psi_t = (t^{1-2v} \chi)_\theta + 2 t^{1-2v} \Phi_\theta\chi + 2\frac{V - v}{t} \Psi, 
\\
& \chi_t =  ( t^{2v-1} \Psi)_\theta. 
\endaligned
\ee
We then consider the initial-value problem with data prescribed on $t=0$ and, more precisely, 
given data $v,\varphi,\psi,q$, we solve \eqref{BCP.10} for $t>0$, with initial data
\be
\label{BCP.30}
\aligned
& V(0,\theta) = v(\theta), \qquad && \Phi(0,\theta) = \varphi(\theta),
\\
&\Psi(0,\theta) = 2v(\theta)\psi(\theta), \qquad && \chi(0,\theta) = q'(\theta).
\endaligned
\ee

%--------------------------------------------------------------------------------------------------------------

\subsection{Heuristics revisited}
\label{HR}
The condition $v\in(0,1)$ 
arises formally from the formulation \eqref{BCP.10} of the Einstein equations. 
Indeed, we expect $V$ to have a finite limit, $v$, on the slice $t=0$. 
However, the first equation in \eqref{BCP.10} is of the form $V_t \sim t^{\overline v}$, 
with $\overline v = \min(2v-1,1-2v)$, unless $\Psi$ or $\chi$ {\sl vanish} on the 
 singularity. This suggests (at least formally) 
 that $V \sim t^{\overline v +1} + V_0$, and according to this, $V$ has a finite limit on $t=0$ if and only if $v \in (0,1)$. 

Thus, if $V$ has a finite limit on the 
coordinate singularity, then this limit must lie in the interval $(0,1)$, 
except possibly on a set of exceptional values where either $\Psi$ or $\chi$ vanish.

Similarly, in view of the third and fourth equations 
we see that if (for instance) $v\ge 1$ at some point, then $\chi_\theta$ must vanish at that point
--in order to ensure that $\Psi$ has a finite limit on the 
singularity. Similarly, $\Psi_\theta$ must vanish where $v\le 0$, to ensure that $\chi$ has a finite limit for $t=0$.

We thus see that, at least formally, the unknowns in \eqref{BCP.10} have finite limits on the
coordinate singularity if either $v \in (0,1)$ or 
\[
\begin{aligned}
&v \ge 1 \text{ and } \chi = \chi_\theta = 0,
\\
&v \le 0 \text{ and } \Psi = \Psi_\theta = 0. 
\end{aligned}
\]
We will see that our numerical results fully agree with this heuristical analysis.

%---------------------------------------------------------------------------------------------------------

\subsection{Pseudo-spectral scheme}

We approximate the Einstein system for the Gow\-dy spacetimes, \eqref{BCP.10}, \eqref{BCP.30}
by using a Fourier pseudo-spectral approximation scheme. We follow the main lines of the schemes developed in Section~\ref{FM}, 
 however some additional care is needed
  for the discretization of the (singular) equations \eqref{BCP.10}.  
Interestingly enough, the heuristics discussed previously provides the necessary insight.

Denote by $\Delta t$ the time-increment and set $t_{n+1/2} := (n+1/2) \Delta t$. 
Consider, for instance, the first equation in  \eqref{BCP.10}. 
Applying the Fourier transform and using the explicit Euler scheme 
for the time derivative, we arrive at the following equations for all $k=-K,\dots,K$ 
\[
\aligned
V^k_{n+1} =& V^k_n -(n+1/2)(ik \Delta t)^2 \Phi_n^k +(n+1/2 ) (ik \Delta t)^2v^k \log t_{n+1/2} 
\\
& - \Big( e^{2\Phi_n}\big( (n+1/2)^{2v-1} (\Delta t)^{2v} \Psi_n^2 - (n+1/2)^{1-2v} (\Delta t)^{2-2v} \chi_n^2 \big) \Big)^k.
\endaligned
\]
The key observation here is that all powers of $\Delta t$ are {\sl positive,} 
at least as long as the velocity is chosen to satisfy $v \in (0,1)$. 
Hence, the above method makes sense under precisely the same conditions we derived by formal analysis.

%======================================================================================================================

\section{Numerical experiments}
\label{NE}

\subsection{Asymptotic velocity inside the interval $(0,1)$}

In this section, we provide extensive numerical tests performed with the methods proposed 
in this paper, and we implement the backward-forward strategy 
sketched earlier in Section~\ref{BM}. 

\begin{figure}[htb]
$$
\includegraphics[width=180pt,keepaspectratio=false]{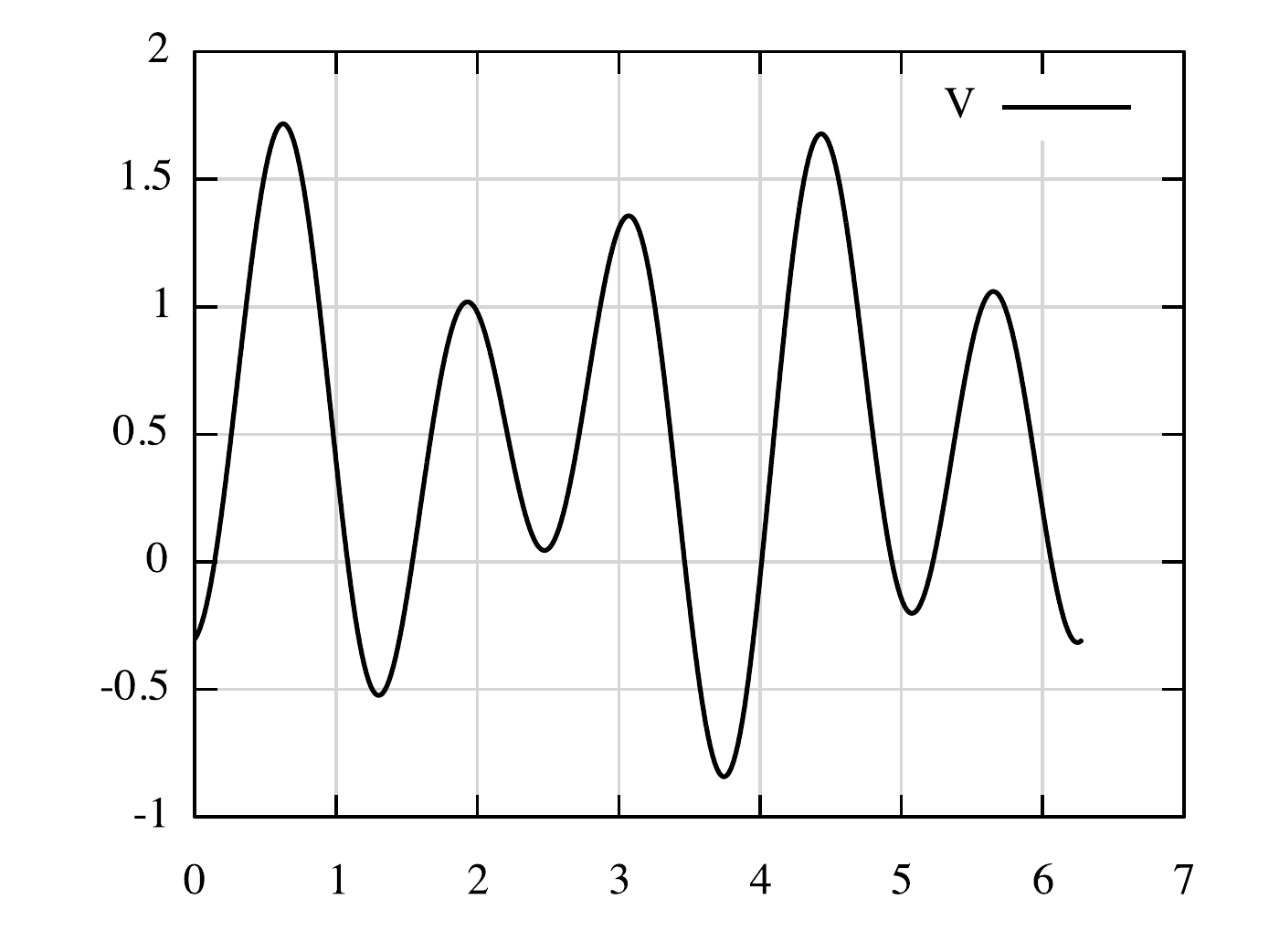}
\includegraphics[width=180pt,keepaspectratio=false]{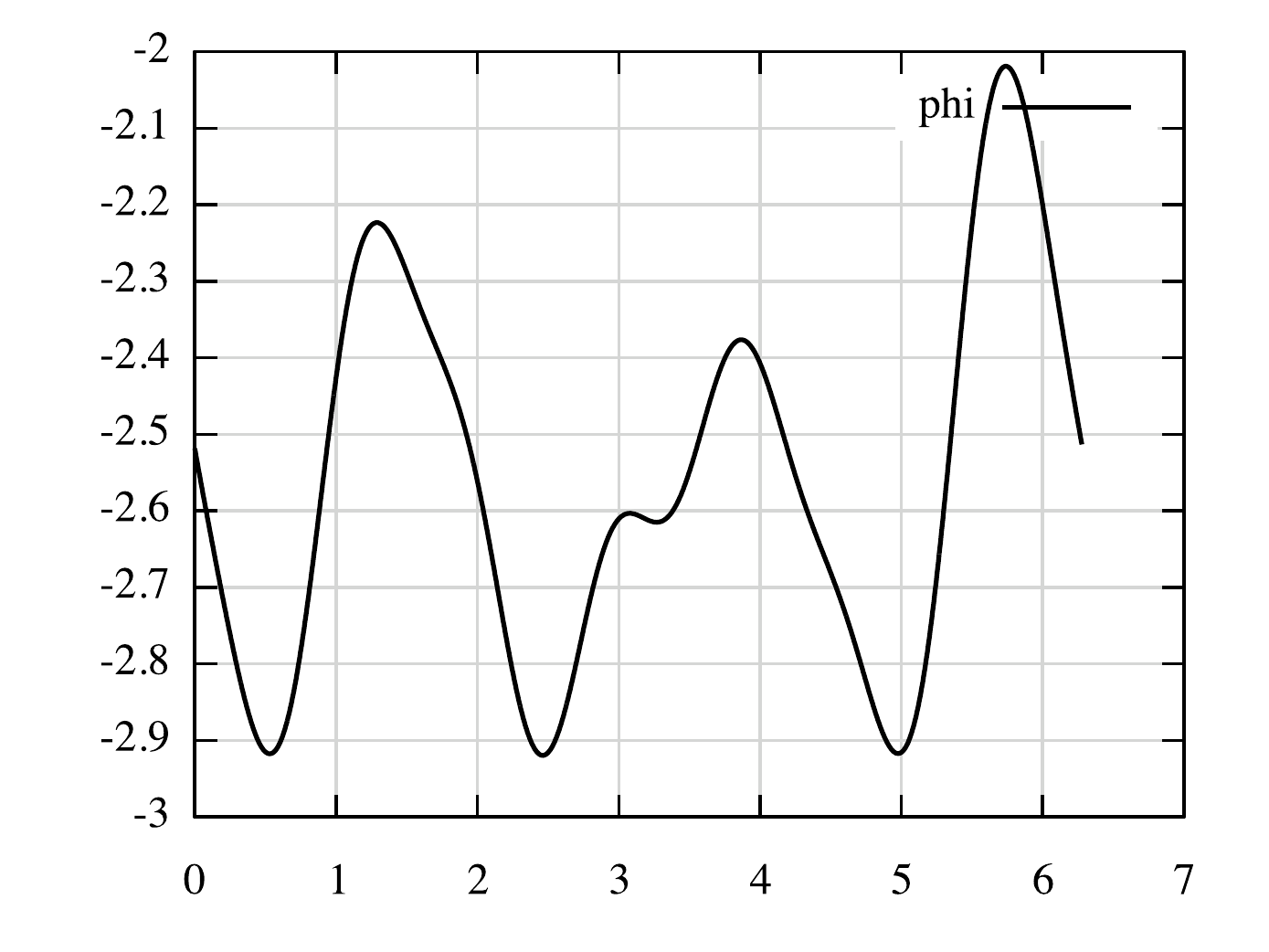}
$$
$$
\includegraphics[width=180pt,keepaspectratio=false]{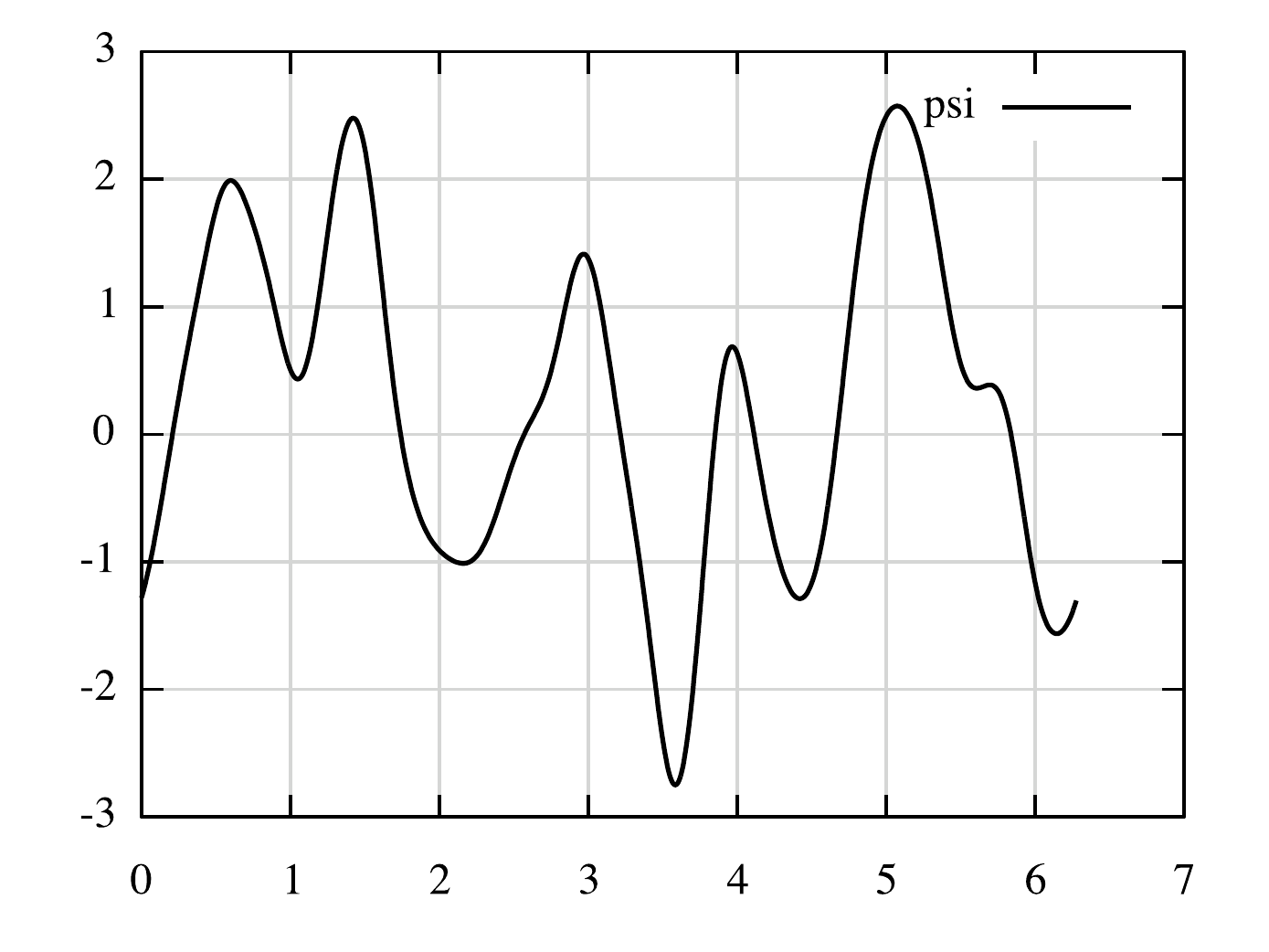}
\includegraphics[width=180pt,keepaspectratio=false]{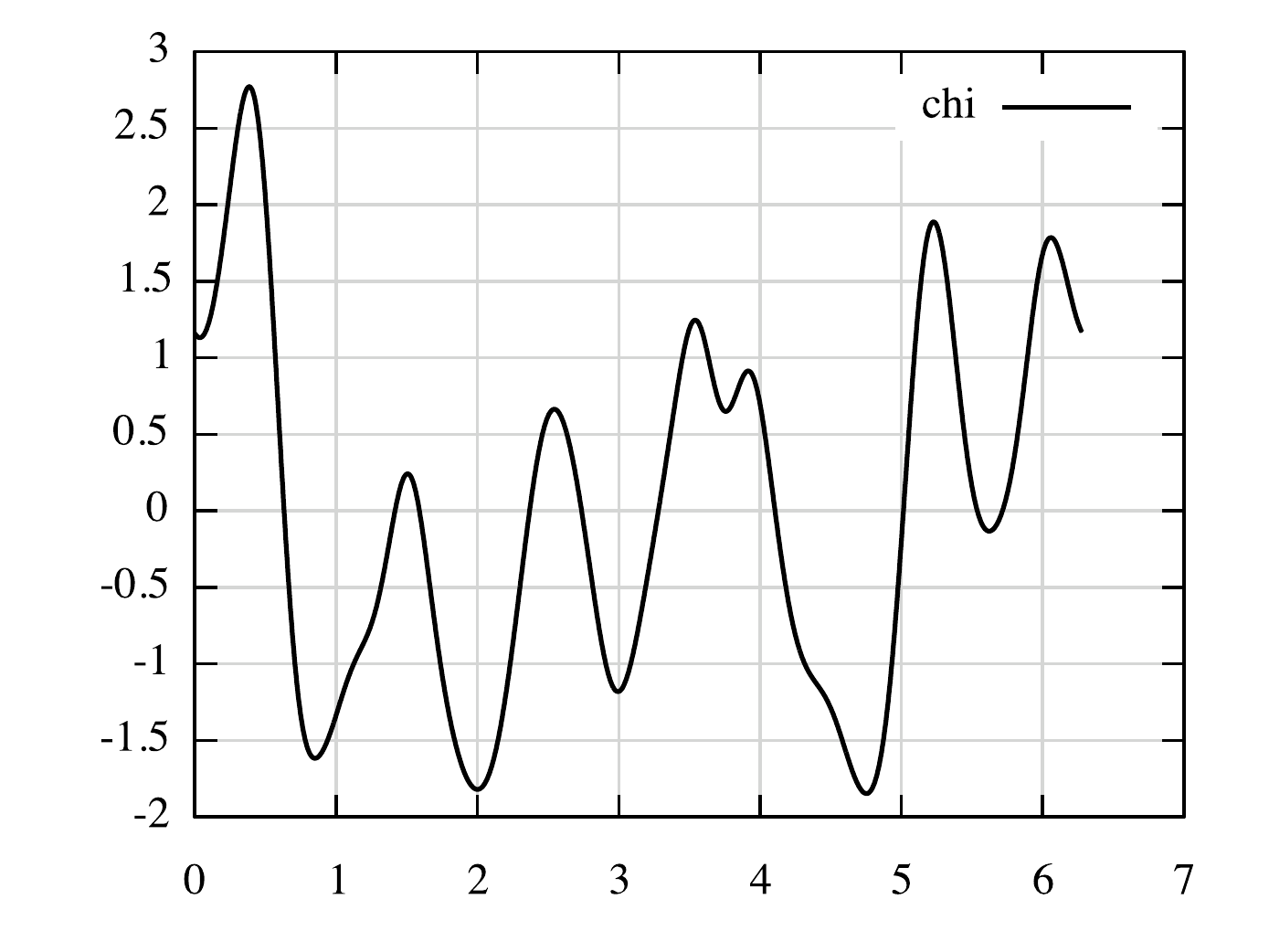}
$$
\caption{Solution computed by the backward code at $t=1$ ($\tau=0$), 
with initial data given by \eqref{BCP.40}.}
% This solution is taken as the initial value for the forward code.}
\label{FIG.100}
\end{figure}

We begin with the case that the asymptotic velocity lies in the interval $(0,1)$. 
We run the backward method described in Section~\ref{BM}, with initial data
\be
\label{BCP.40}
\aligned
&V(0) =0.2 \sin 3\theta + 0.1 \cos 5\theta+0.5, 
\\
& \Phi(0)=0.35 \sin 4 \theta +0.1 \cos 1 \theta +0.2 \sin 7 \theta +0.3 \cos 5 \theta +0.5 \sin 3 \theta + C,
\\
& \Psi(0)=0.25 \sin 4 \theta +0.5 \cos 3 \theta +0.4 \sin 2 \theta +0.2 \cos 6 \theta +0.15 \sin 7 \theta,
\\
 &\chi(0)=- \sin 2 \theta,
 \endaligned
\ee
where the constant $C$ is chosen so that the constraint \eqref{GS.Einstein3} holds.
In Figure~\ref{FIG.100}, we display the solution computed by the backward code at the time $t=1$.

\begin{figure}[htbp]
$$
\includegraphics[width=180pt,keepaspectratio=false]{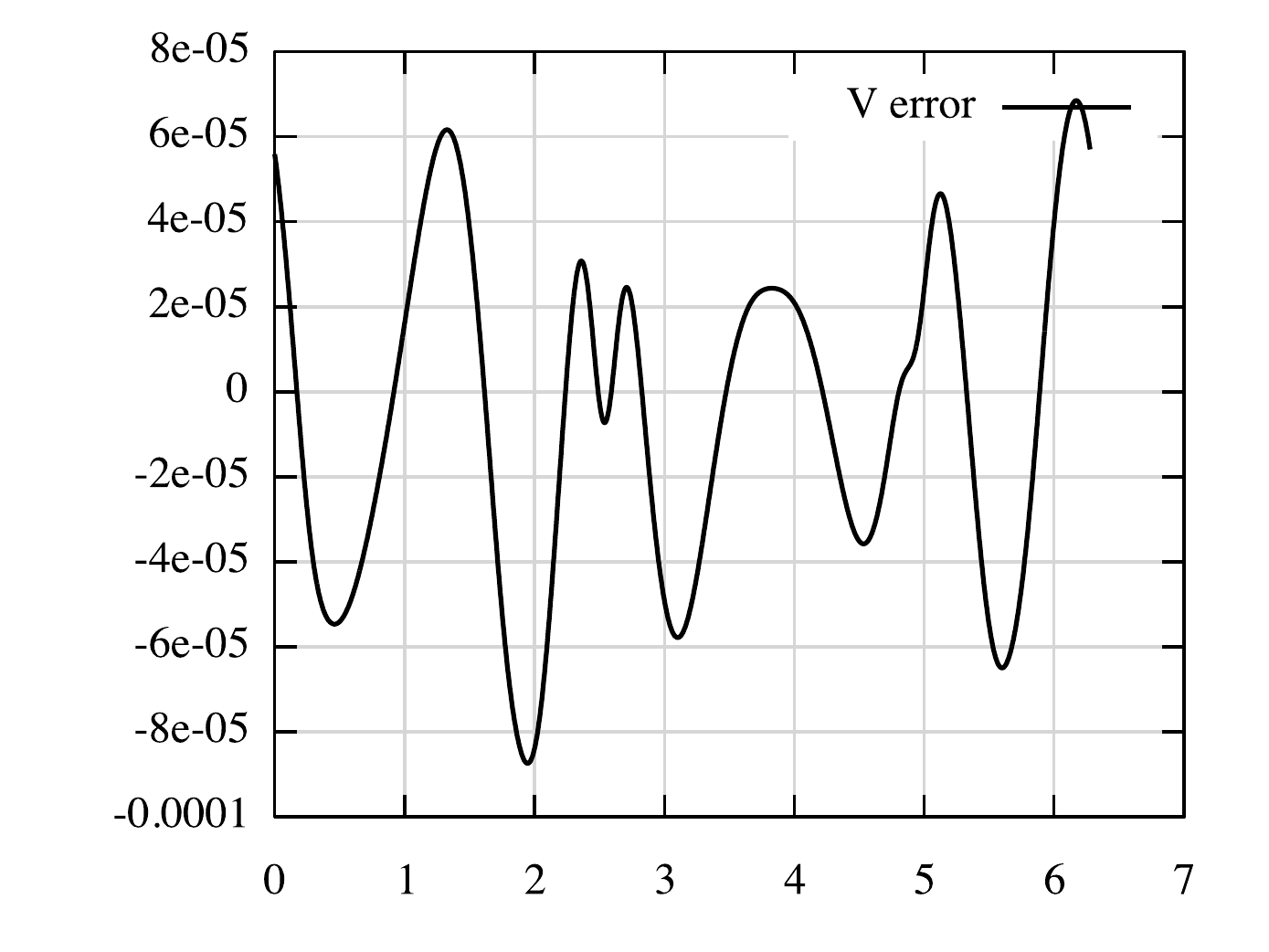}
\includegraphics[width=180pt,keepaspectratio=false]{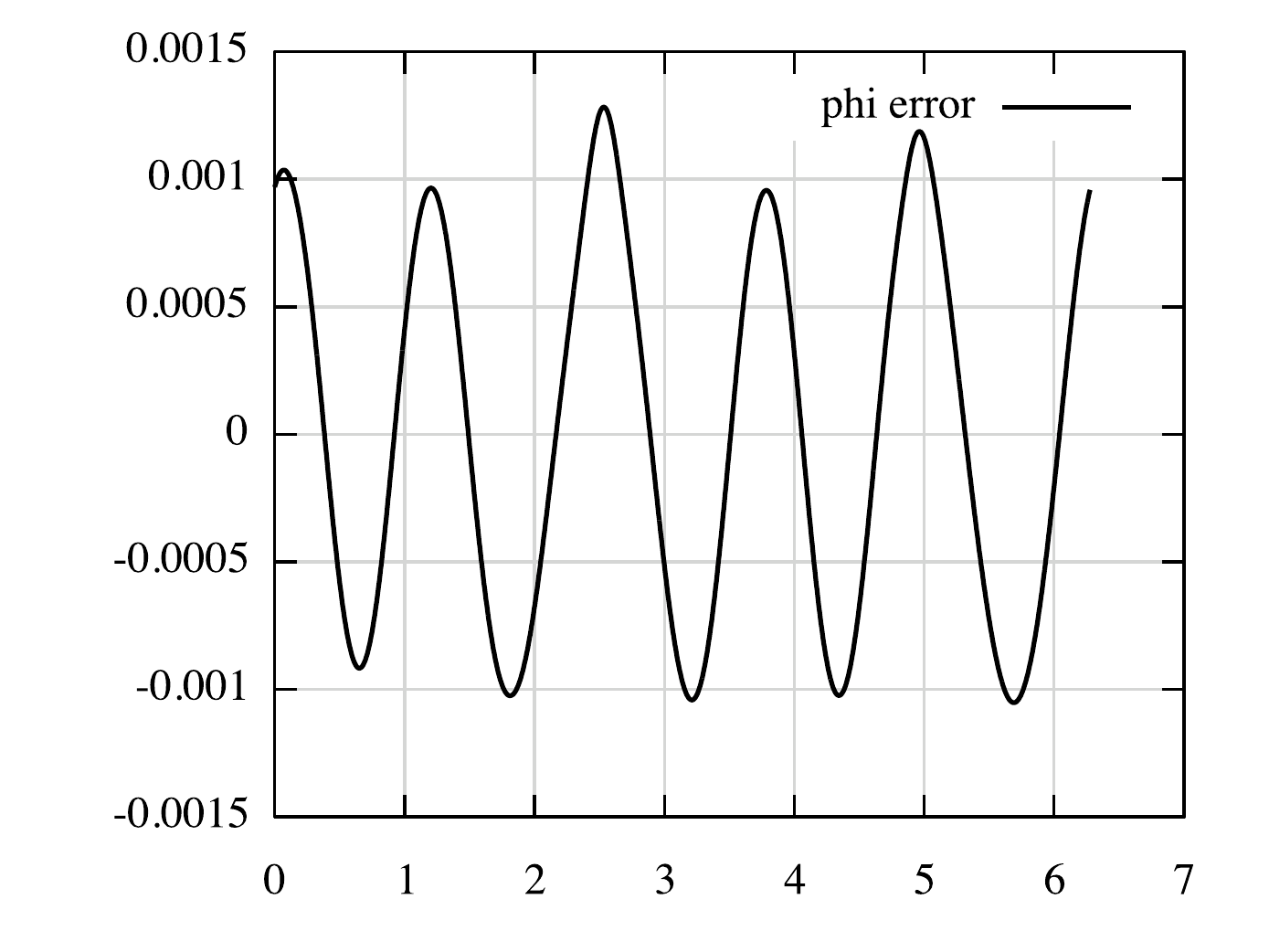}
$$
$$
\includegraphics[width=180pt,keepaspectratio=false]{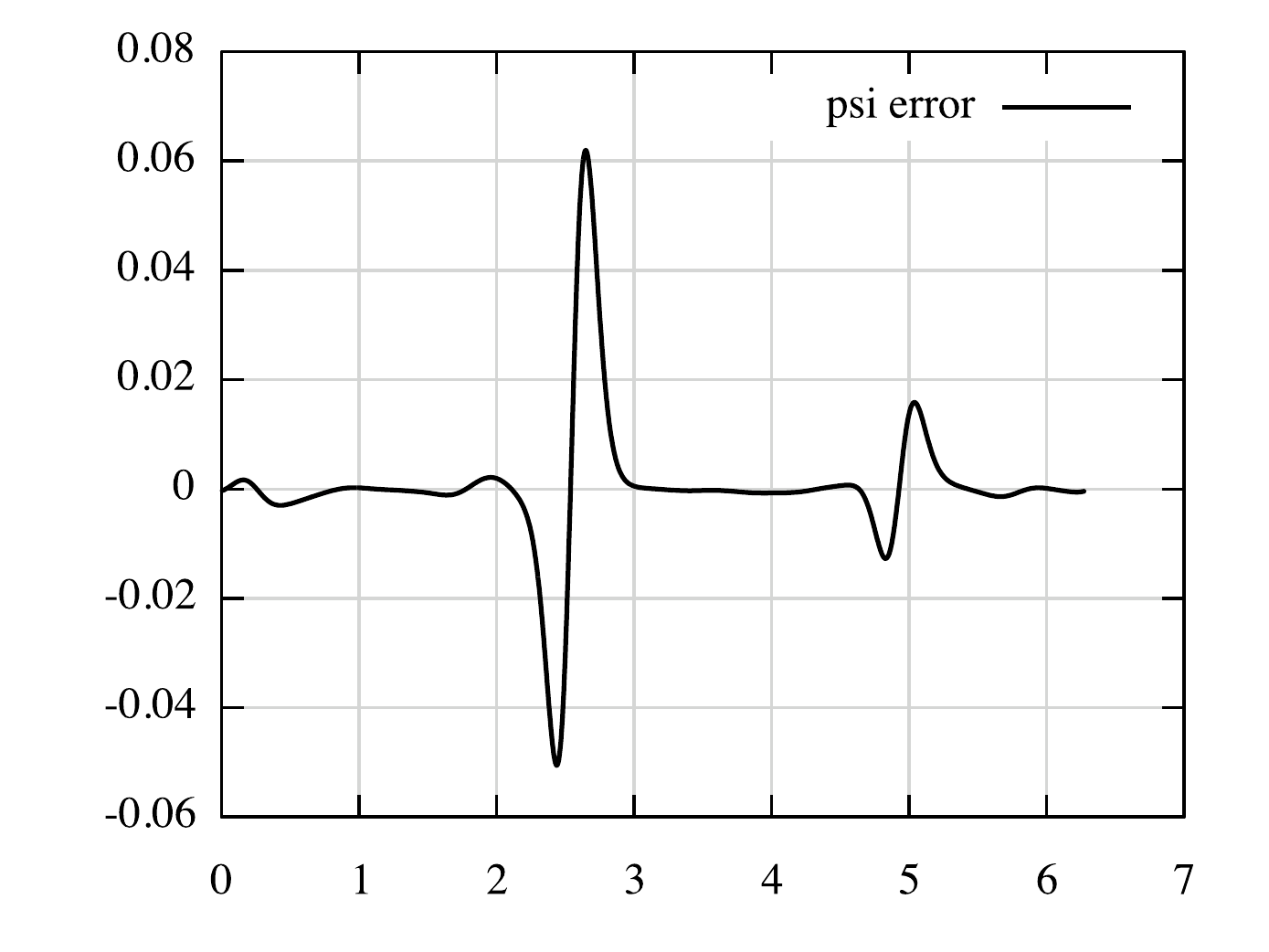}
\includegraphics[width=180pt,keepaspectratio=false]{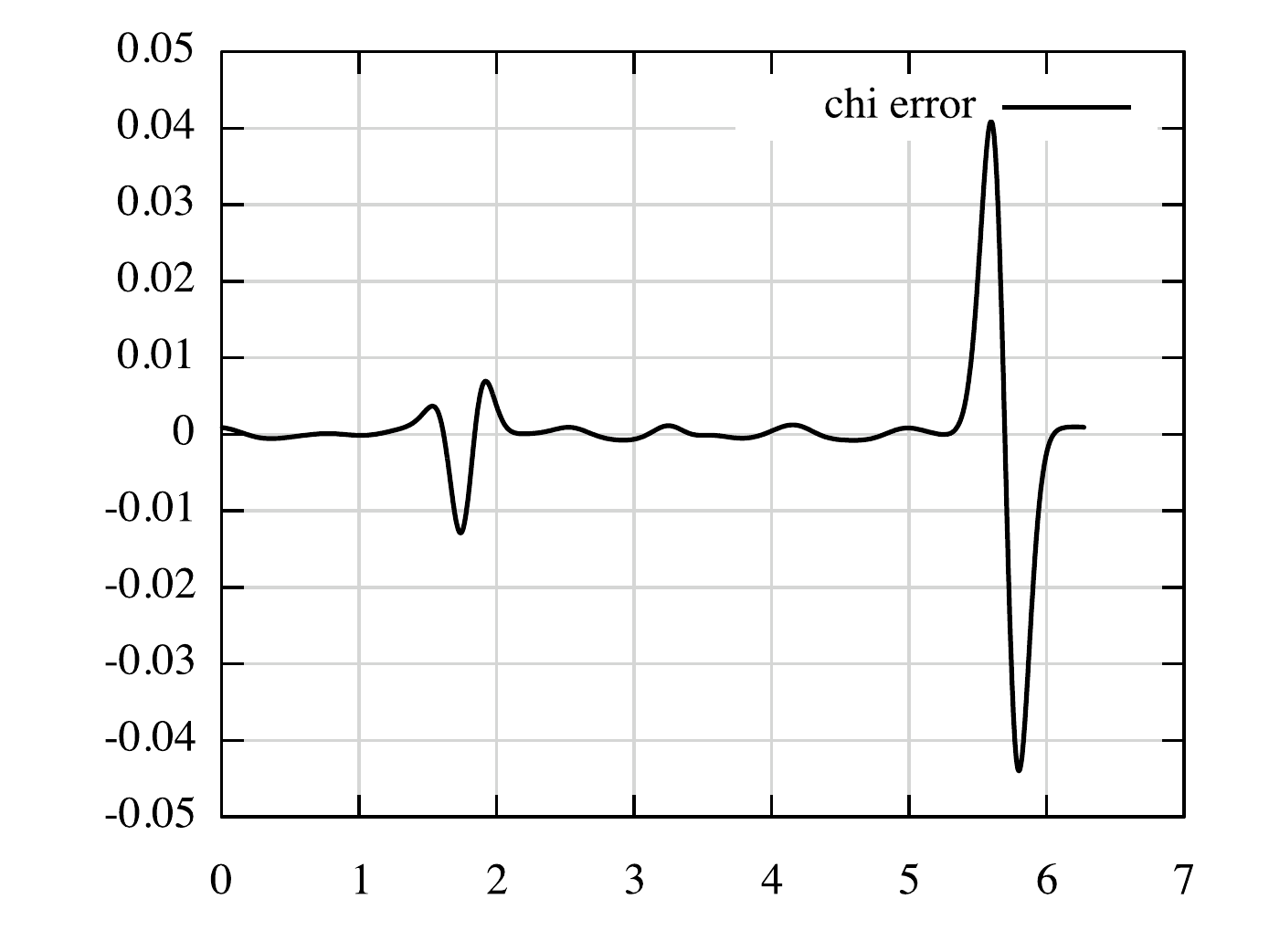}
$$
\caption{Error between the first iteration of the backward method 
and the numerical values after backward and forward evolutions, $\tau=-\log(\Delta t/2)$.}  
%with $\Delta t$ the time increment of the backward method.}
\label{FIG.150}
\end{figure}

Next, as outlined in the previous section, we use the above result as an initial data at $\tau=0$ 
for the forward method. Finally, at the time $\tau =-\log(\Delta t/2)$, we 
can compare the numerical asymptotic velocity $V$ with 
the first iteration of the backward code. At $\tau = -\log(\Delta t/2) +1$, we can also 
compare the numerical asymptotic velocity $V$ with the initial data originally prescribed in
the backward code. We proceed similarly for the other unknowns. 
These last two results have been found to be completely similar, and so we plot the corresponding error at 
the time $\tau=-\log(\Delta t/2)+1$, only. See Figure~\ref{FIG.150}. 

\begin{figure}[htbp]
$$
\includegraphics[width=180pt,keepaspectratio=false]{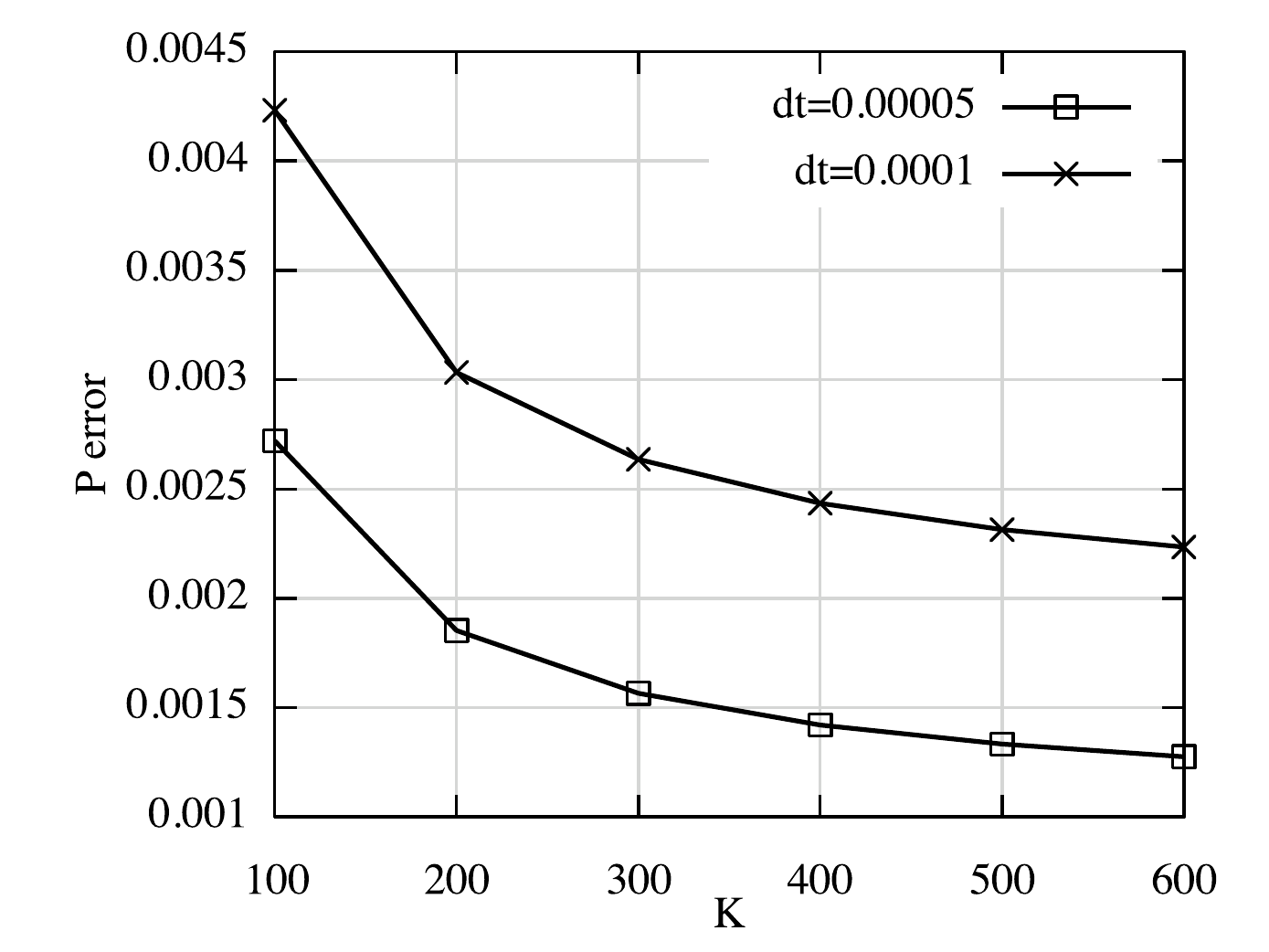}
\includegraphics[width=180pt,keepaspectratio=false]{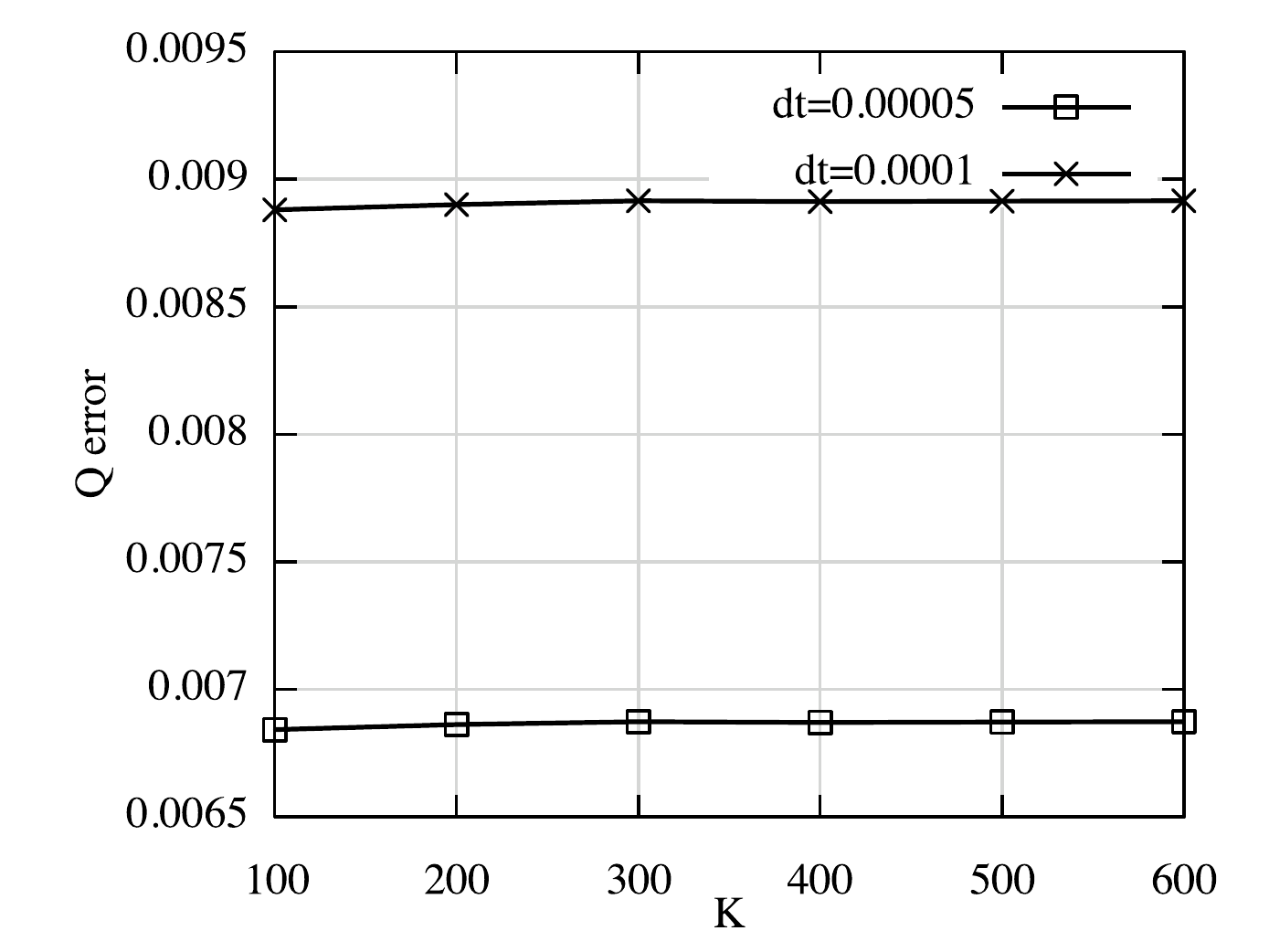}
$$
$$
\includegraphics[width=180pt,keepaspectratio=false]{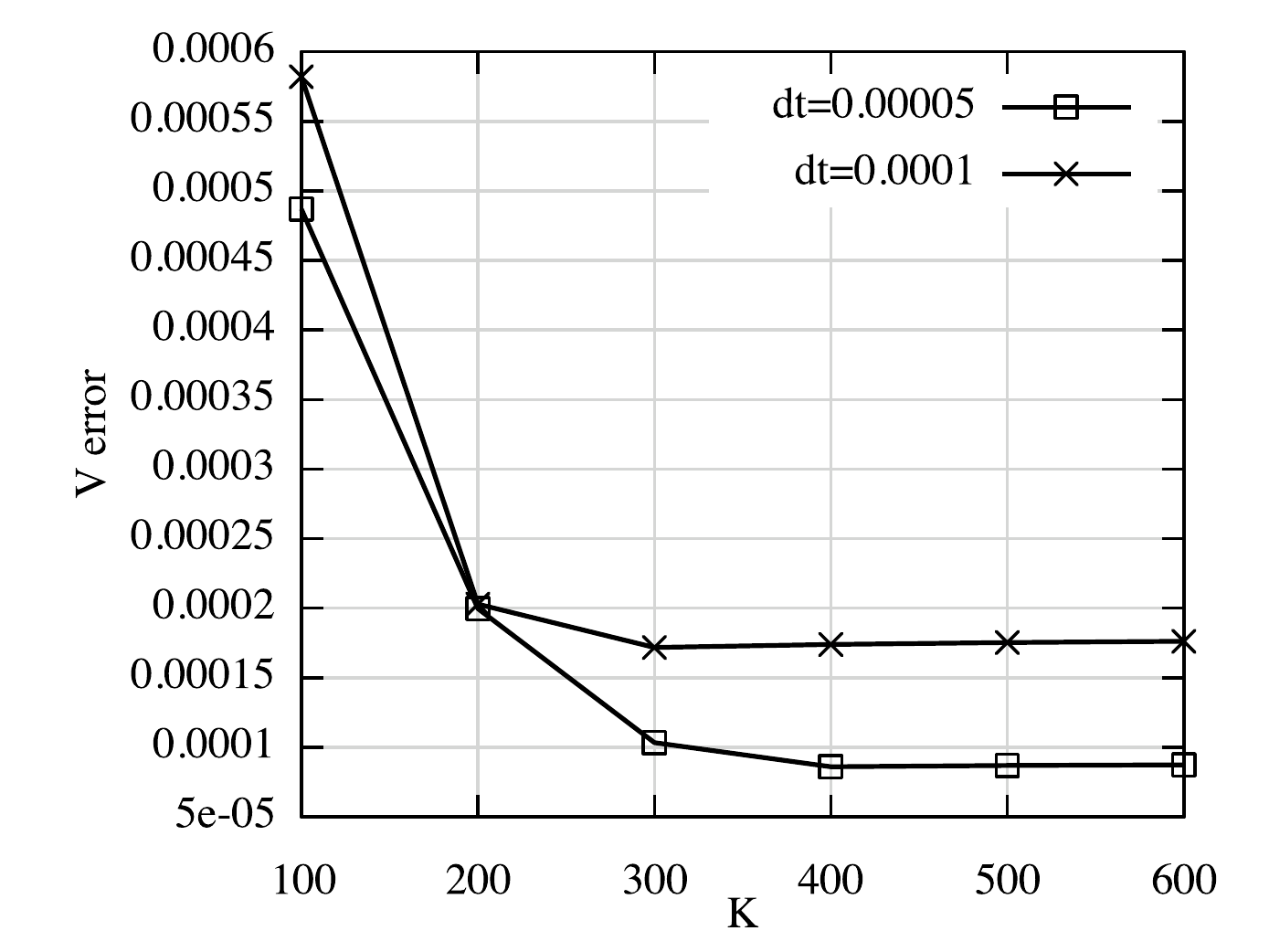}
\includegraphics[width=180pt,keepaspectratio=false]{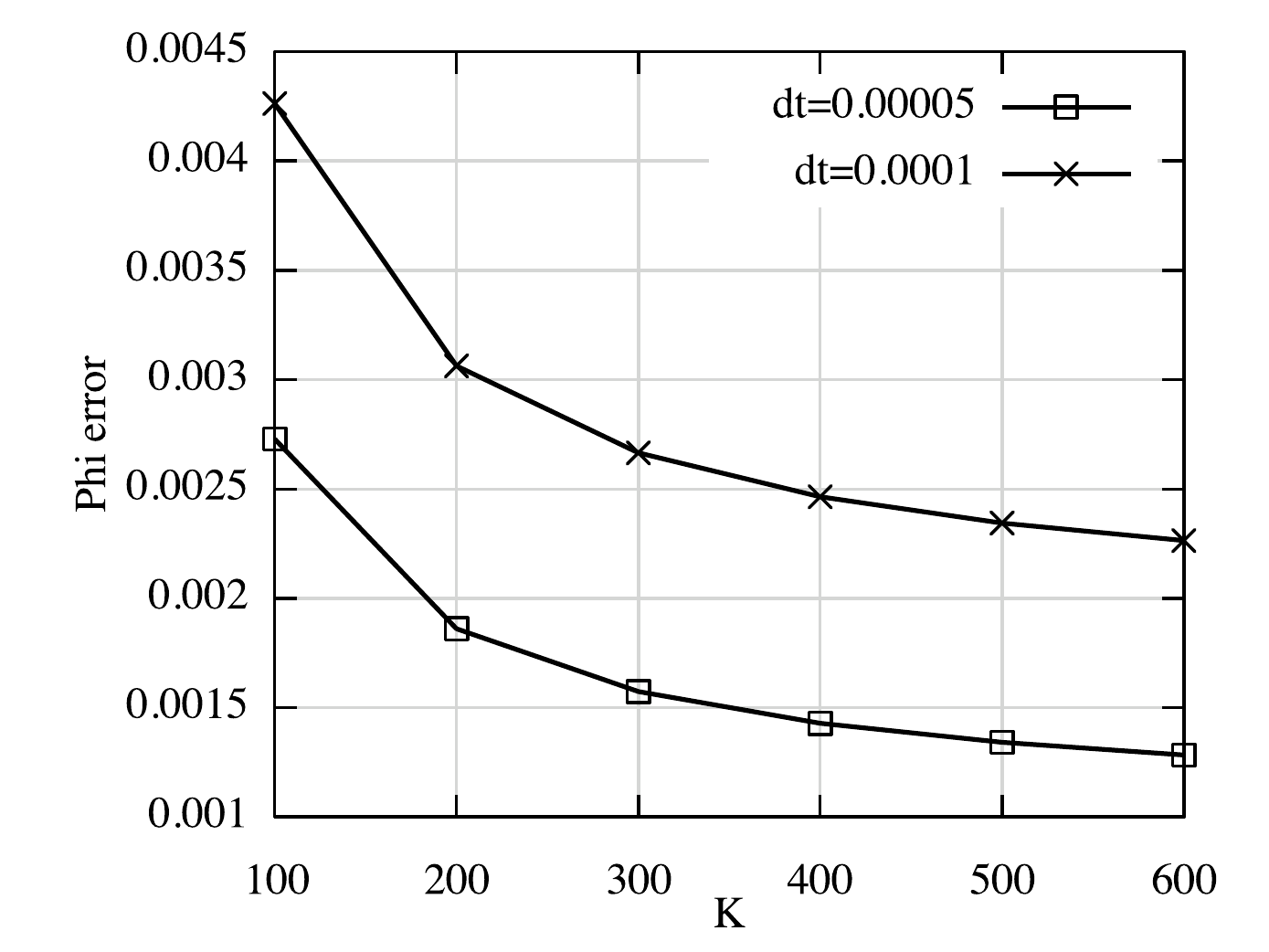}
$$
$$
\includegraphics[width=180pt,keepaspectratio=false]{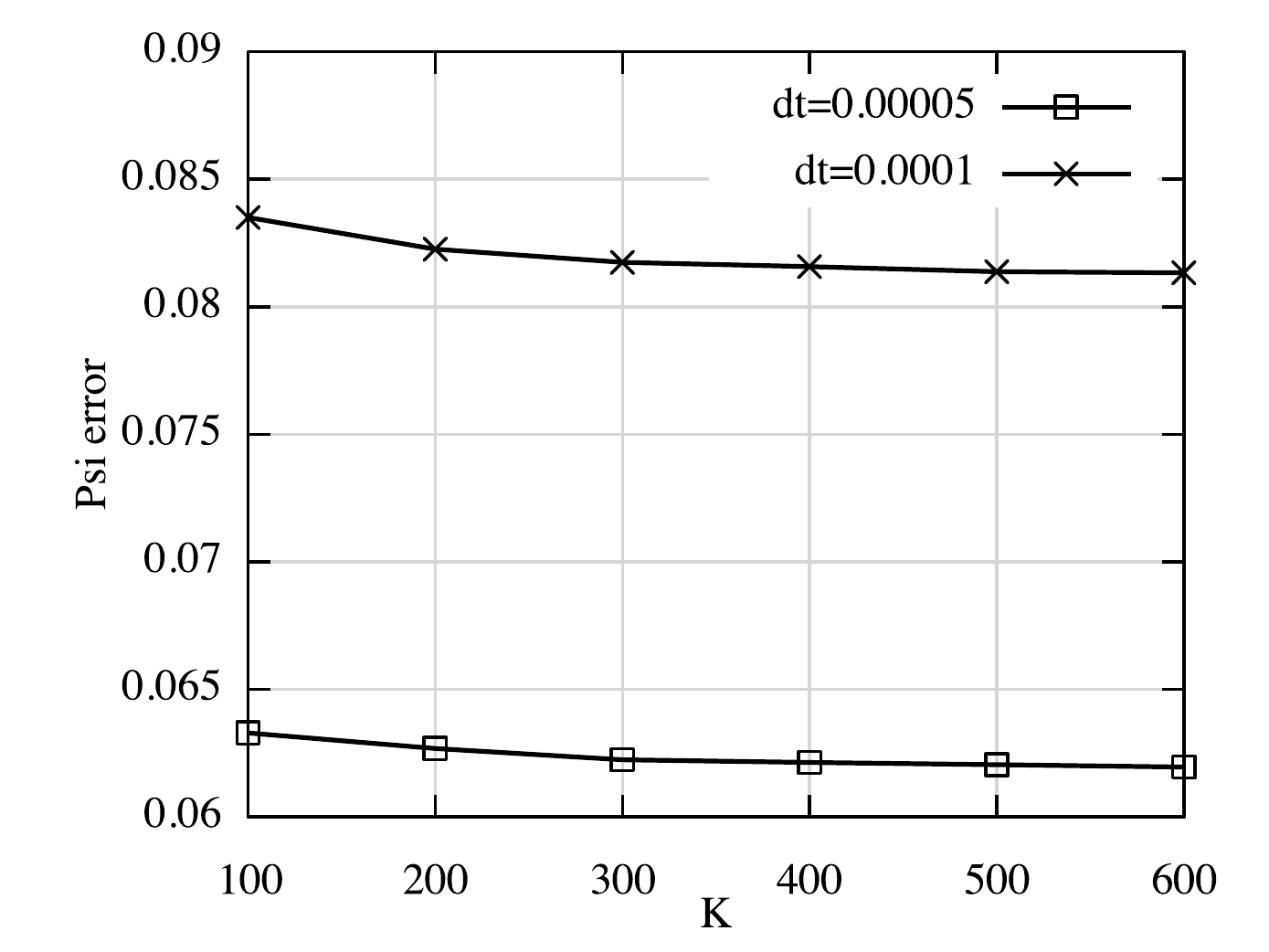}
\includegraphics[width=180pt,keepaspectratio=false]{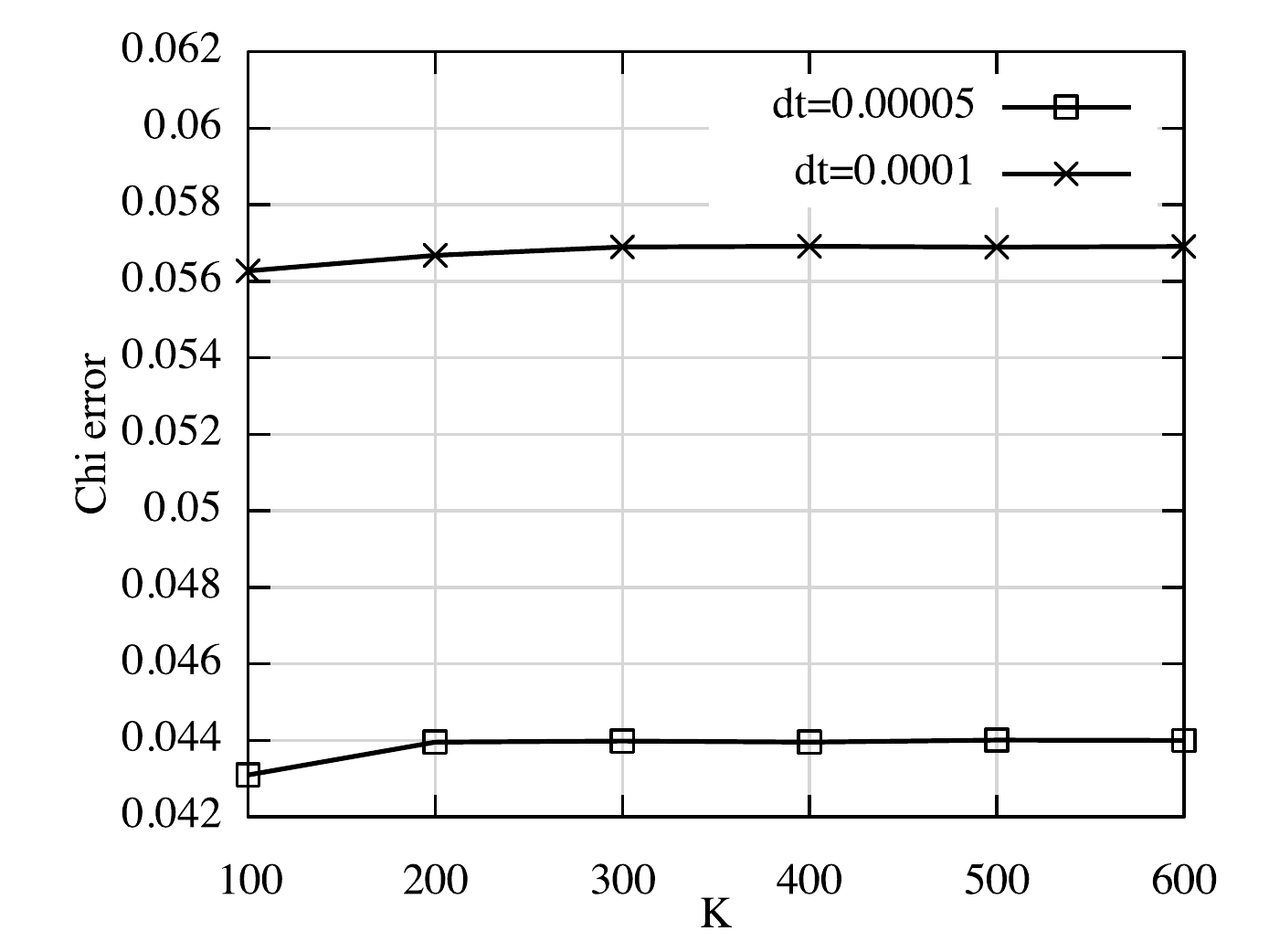}
$$
\caption{Error between the backward-forward solution and the first iteration of the backward code 
at $\tau=-\log(\Delta t/2)$, as a function of the number of points $K$ (for different time steps).}
\label{FIG.350}
\end{figure}

\begin{figure}[htbp]
$$
\includegraphics[width=180pt,keepaspectratio=false]{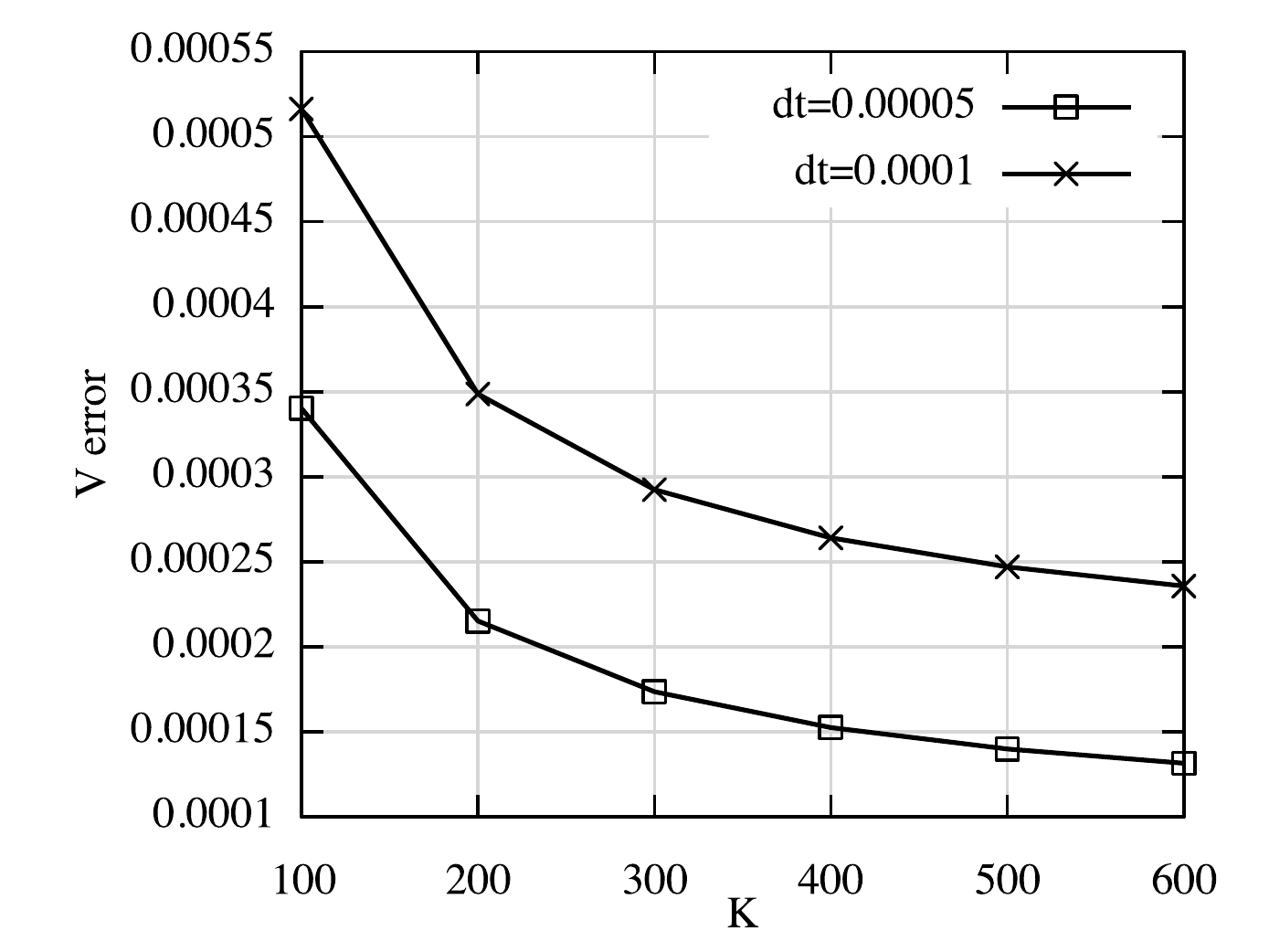}
\includegraphics[width=180pt,keepaspectratio=false]{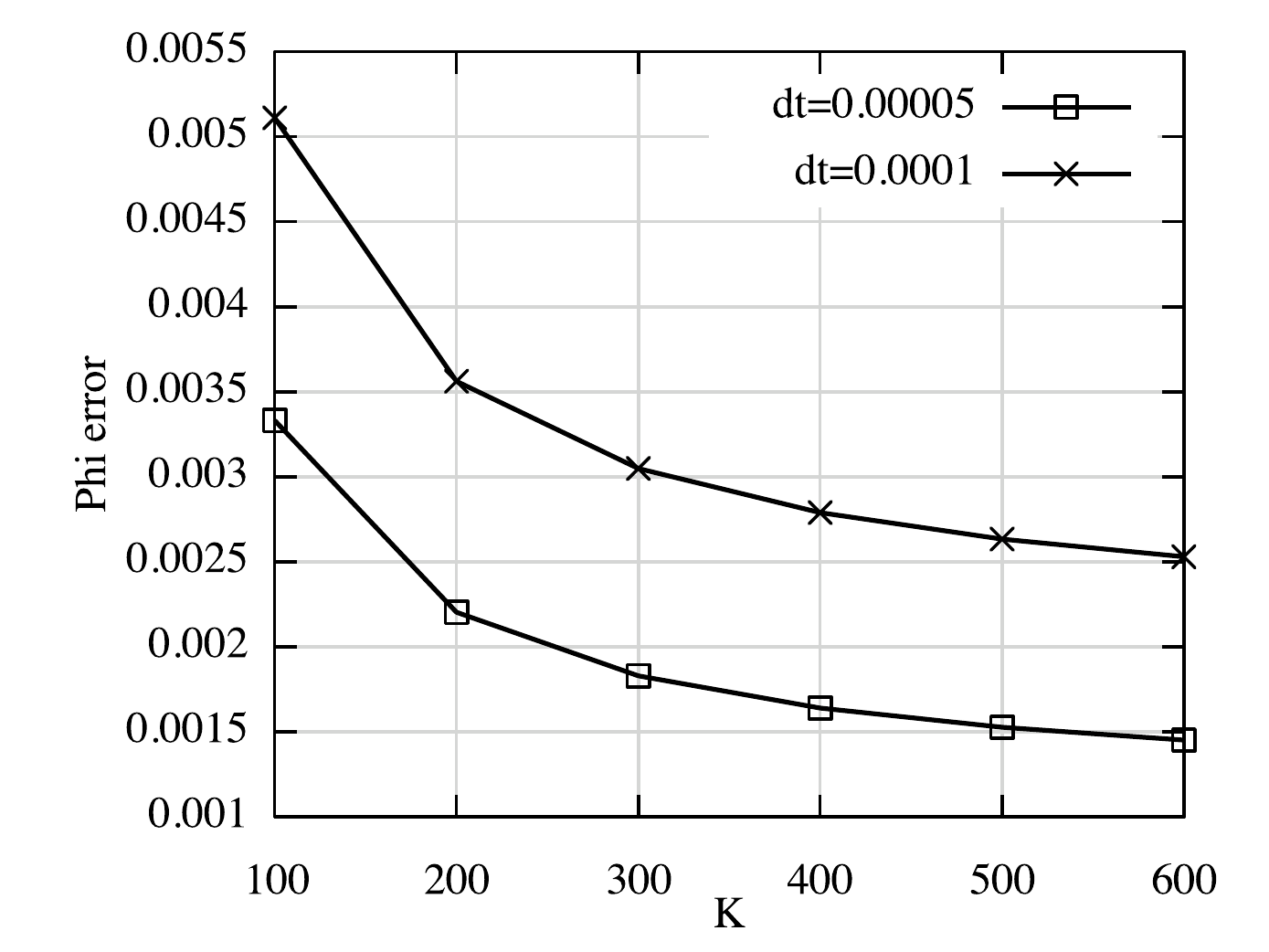}
$$
$$
\includegraphics[width=180pt,keepaspectratio=false]{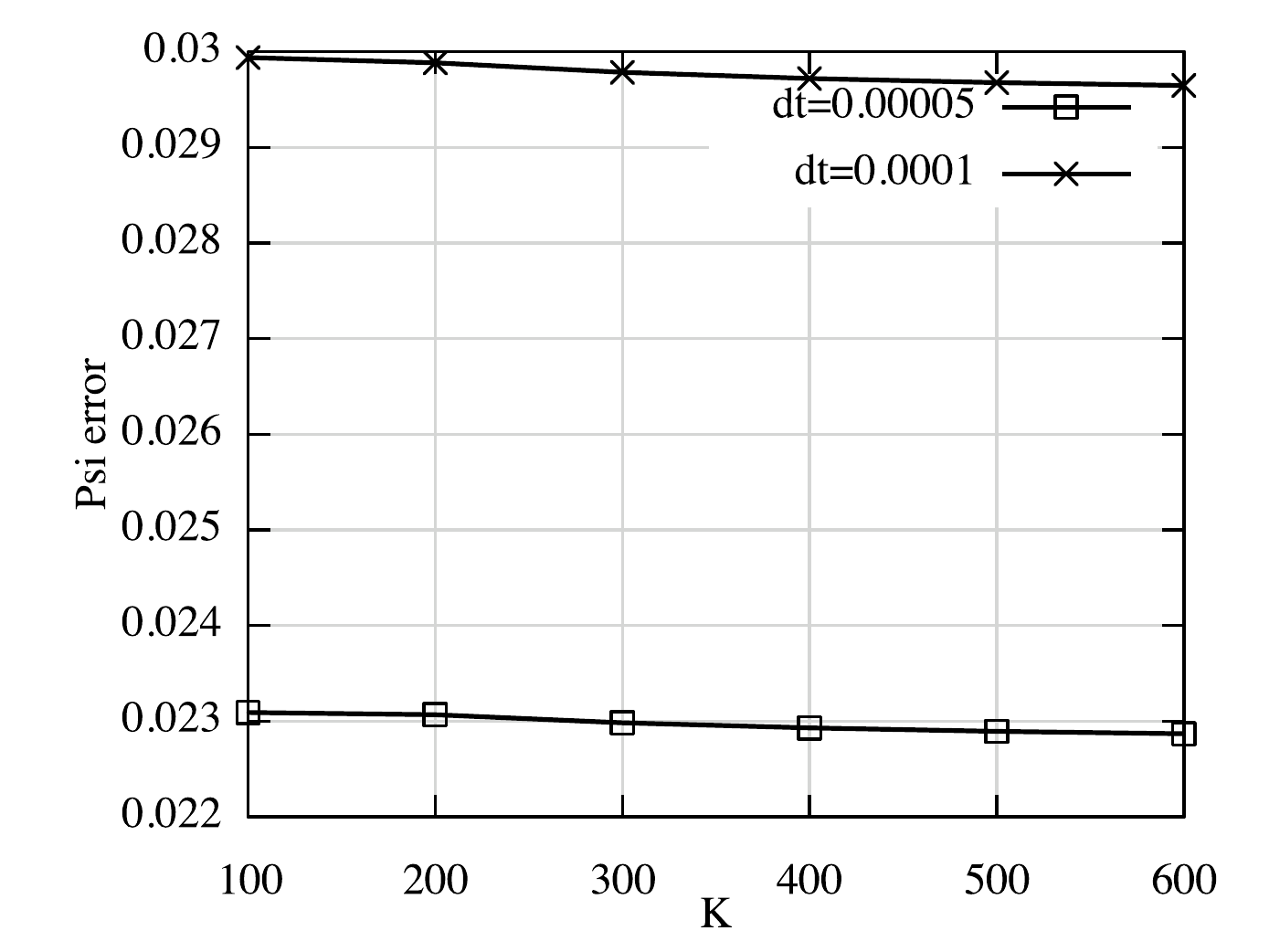}
\includegraphics[width=180pt,keepaspectratio=false]{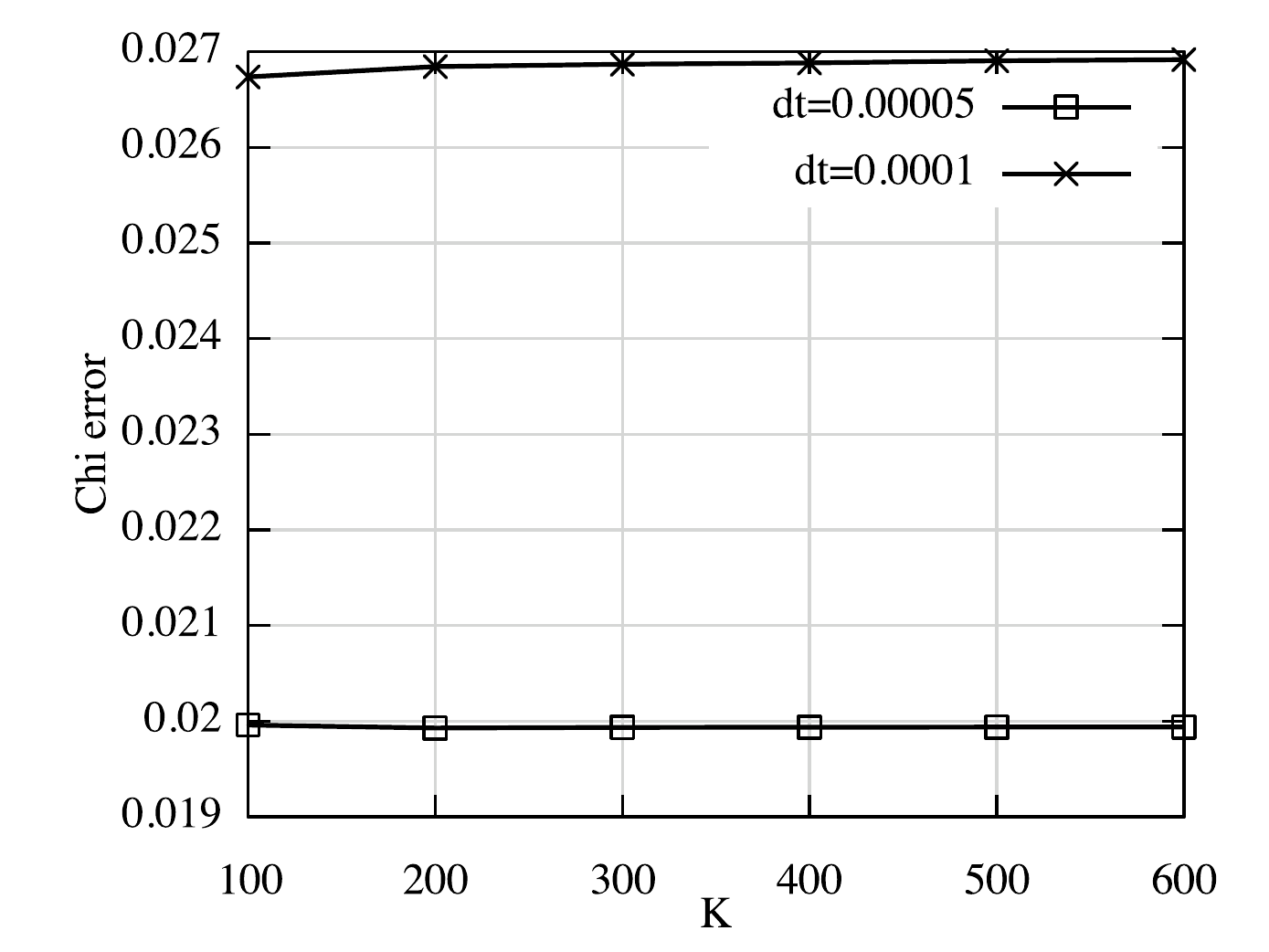}
$$
\caption{Error between the backward-forward solution and the initial data of the backward code at $\tau=-\log(\Delta t/2)+1$, 
as a function of the number of points $K$ (for different time steps).}
\label{FIG.355}
\end{figure}

In Figures~\ref{FIG.350} and \ref{FIG.355} we plot the maximum value of this error as a function of the 
discretization parameter $K$ (recall that the total number of points is $2K+1$) for various choice of time steps. 

\begin{figure}
$$
\includegraphics[width=180pt,keepaspectratio=false]{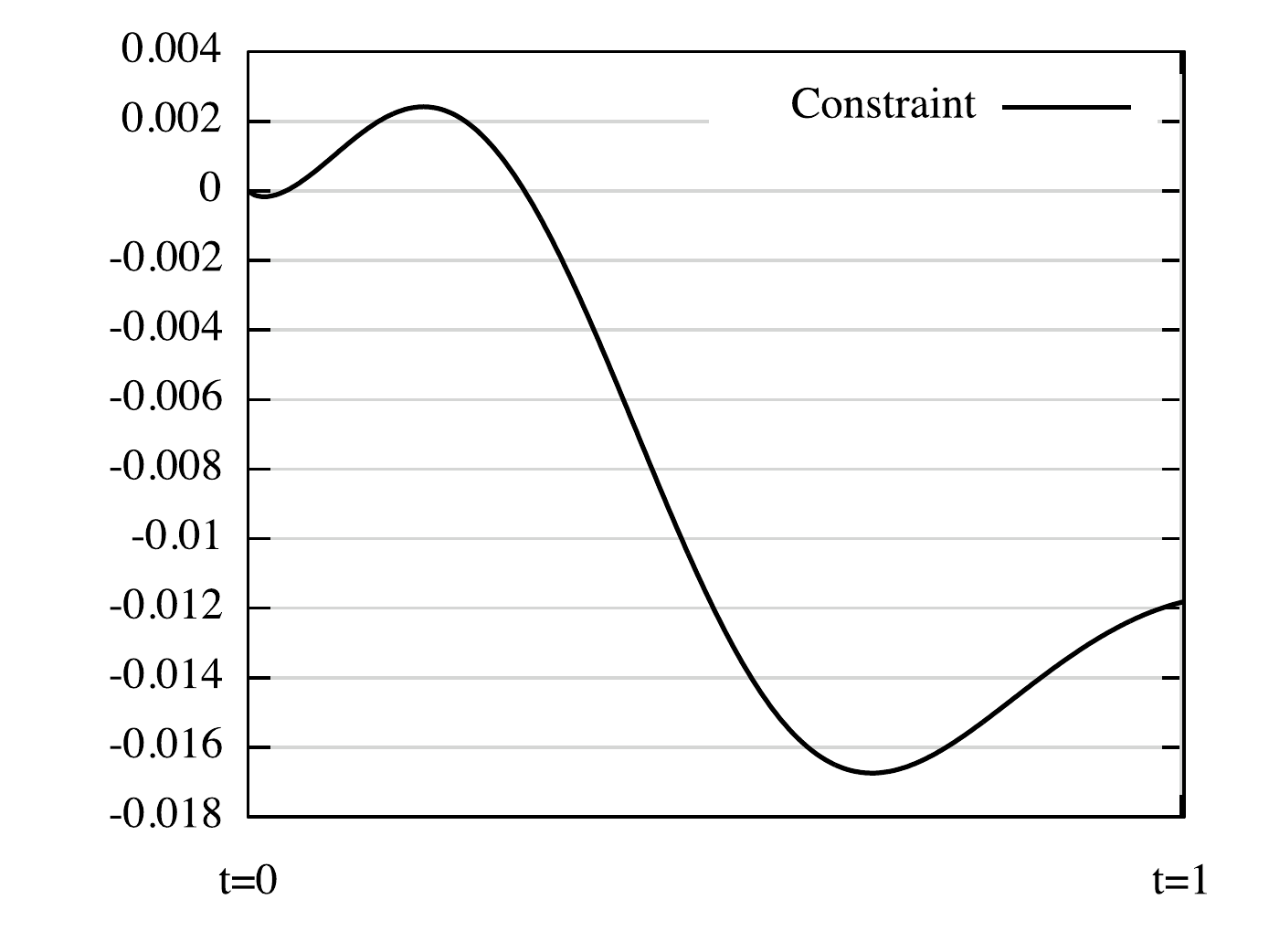}
\includegraphics[width=180pt,keepaspectratio=false]{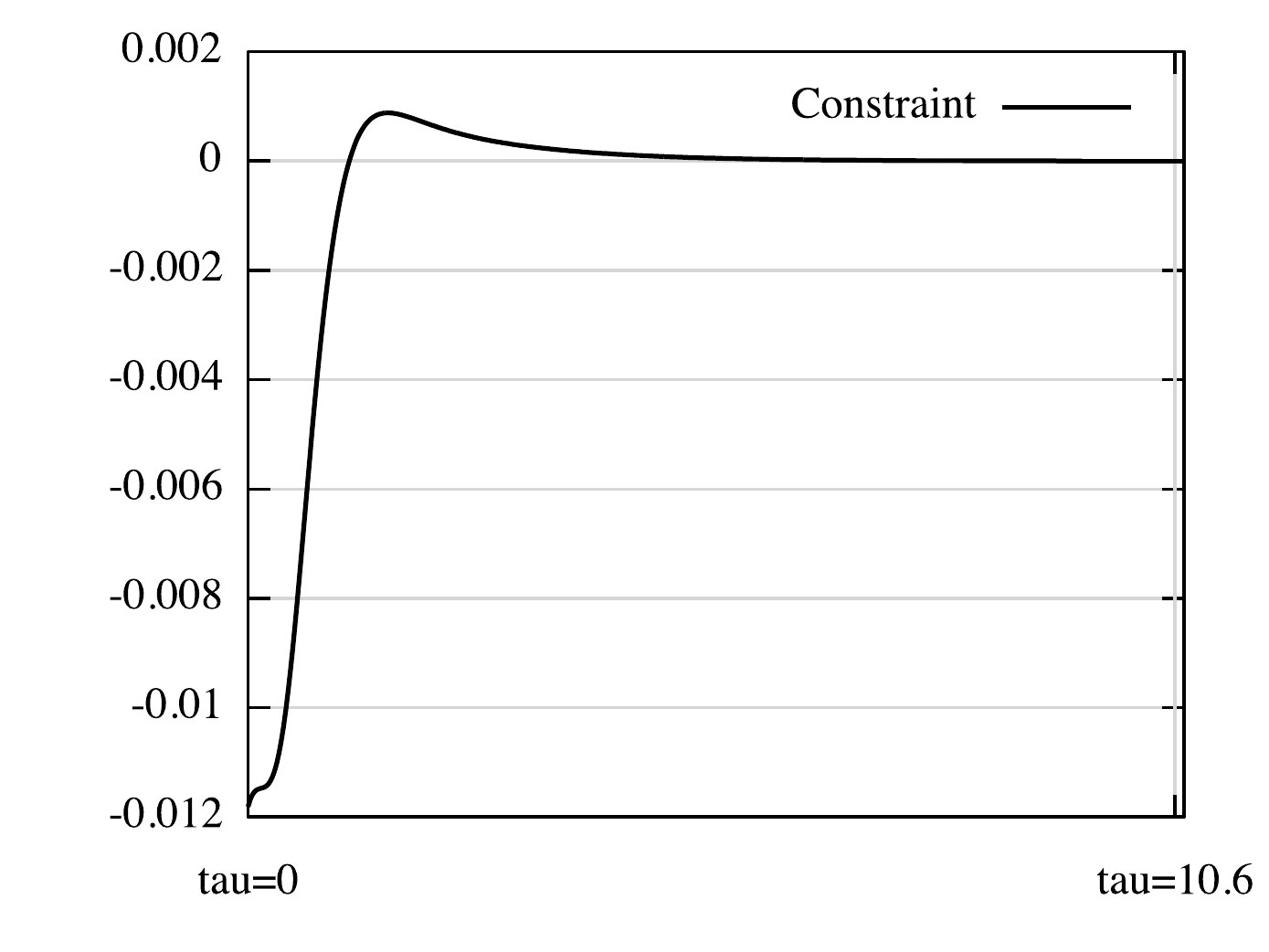}
$$
\caption{Evolution of the constraint \eqref{GS.Einstein3} during the backward evolution (left) and the forward evolution (right).}
\label{FIG.357}
\end{figure}

In Figure \ref{FIG.357}, we show the evolution of the constraint \eqref{GS.Einstein3} during the backward and the forward evolutions.

%--------------------------------------------------------------------------------------------------------------------

\subsection{Asymptotic velocity outside the interval $(0,1)$}

\subsubsection{Agreement with previous numerical results}

We now use our numerical methods to investigate the exceptional situation 
in which the asymptotic velocity $v$ takes some values outside the interval $(0,1)$. 
As observed in the heuristic discussion above, this may only occur at those points where 
either $\Psi$ or $\chi$ vanish on the 
singularity. Furthermore, to ensure that $\Psi$ and $\chi$ approach to finite and regular limits, 
we must assume that $\chi_\theta$ and $\Psi_\theta$ also vanish at those points.

\begin{figure}
$$
\includegraphics[width=180pt,keepaspectratio=false]{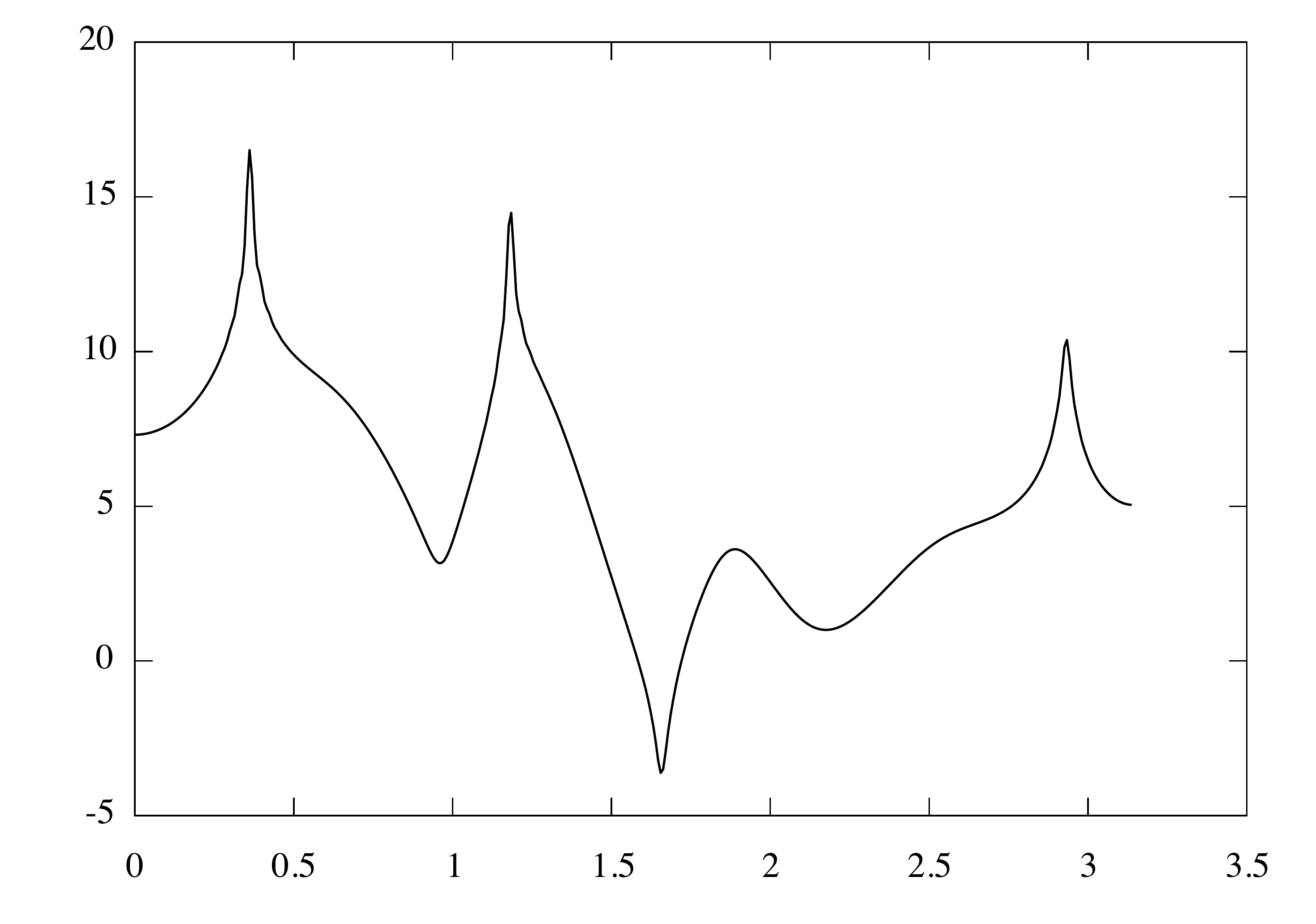}
\includegraphics[width=180pt,keepaspectratio=false]{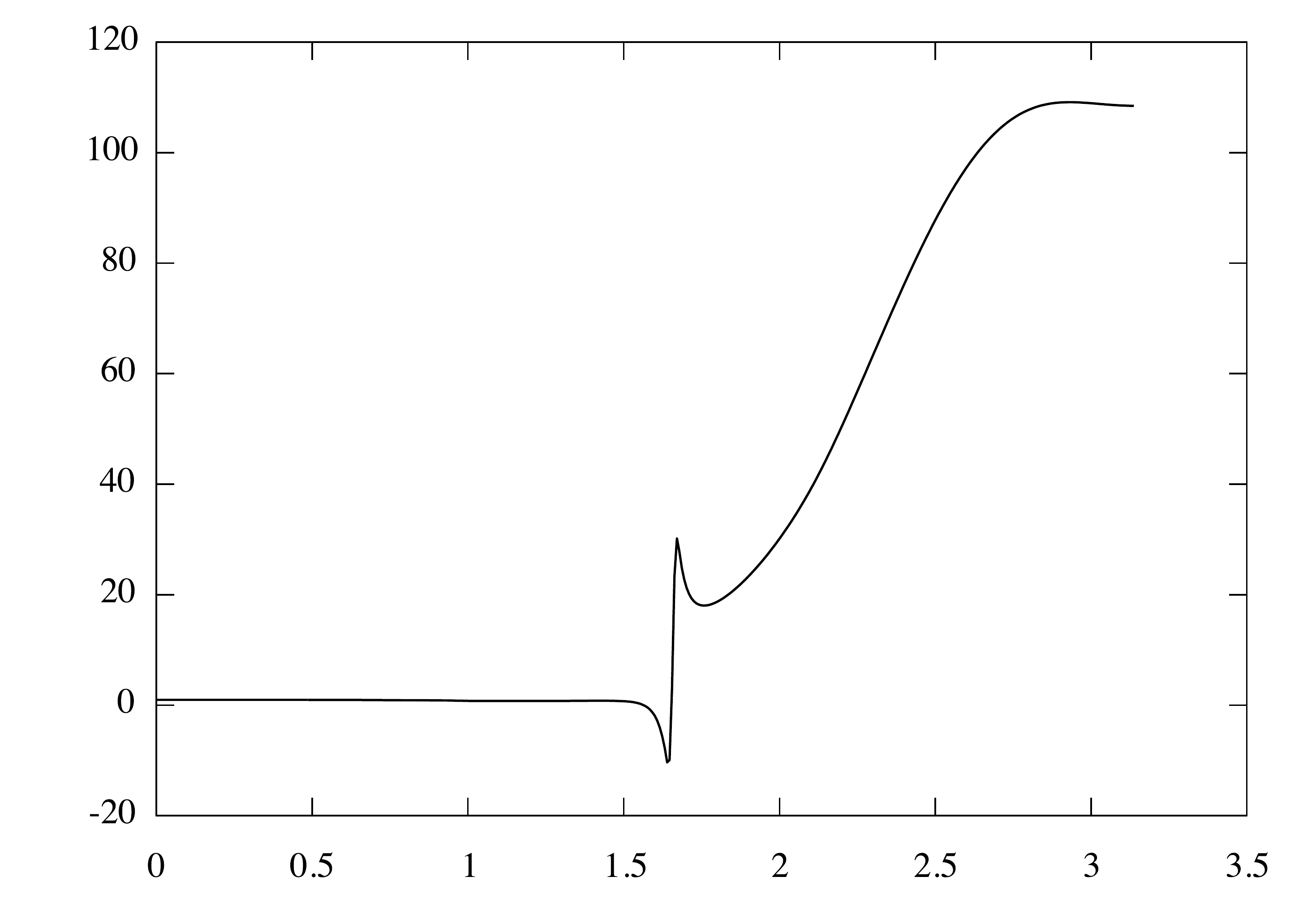}
$$
\caption{$P$ (left) and $Q$ (right) at $\tau=10$, as computed by the forward code for initial data producing spikes.}
\label{FIG.450}
\end{figure}

Figure \ref{FIG.450} displays the metric coefficients $P,Q$ computed by
the forward code for the (large) initial data
\be
\label{BCP.42}
\aligned
&V(0) = 5 \cos \theta, \qquad X(0) = -\sin \theta, \qquad Y(0) = W(0) = 0,
\endaligned
\ee
which are known to produce spikes in the velocity (and, hence, also in $P$ and $Q$). Our results are in complete 
agreement with the previous work \cite{Garfinkle}.

%-------------------------------------------------------------------------------------

\subsubsection{The case that $\chi$ vanishes on an interval}

\begin{figure}[htbp]
$$
\includegraphics[width=180pt,keepaspectratio=false]{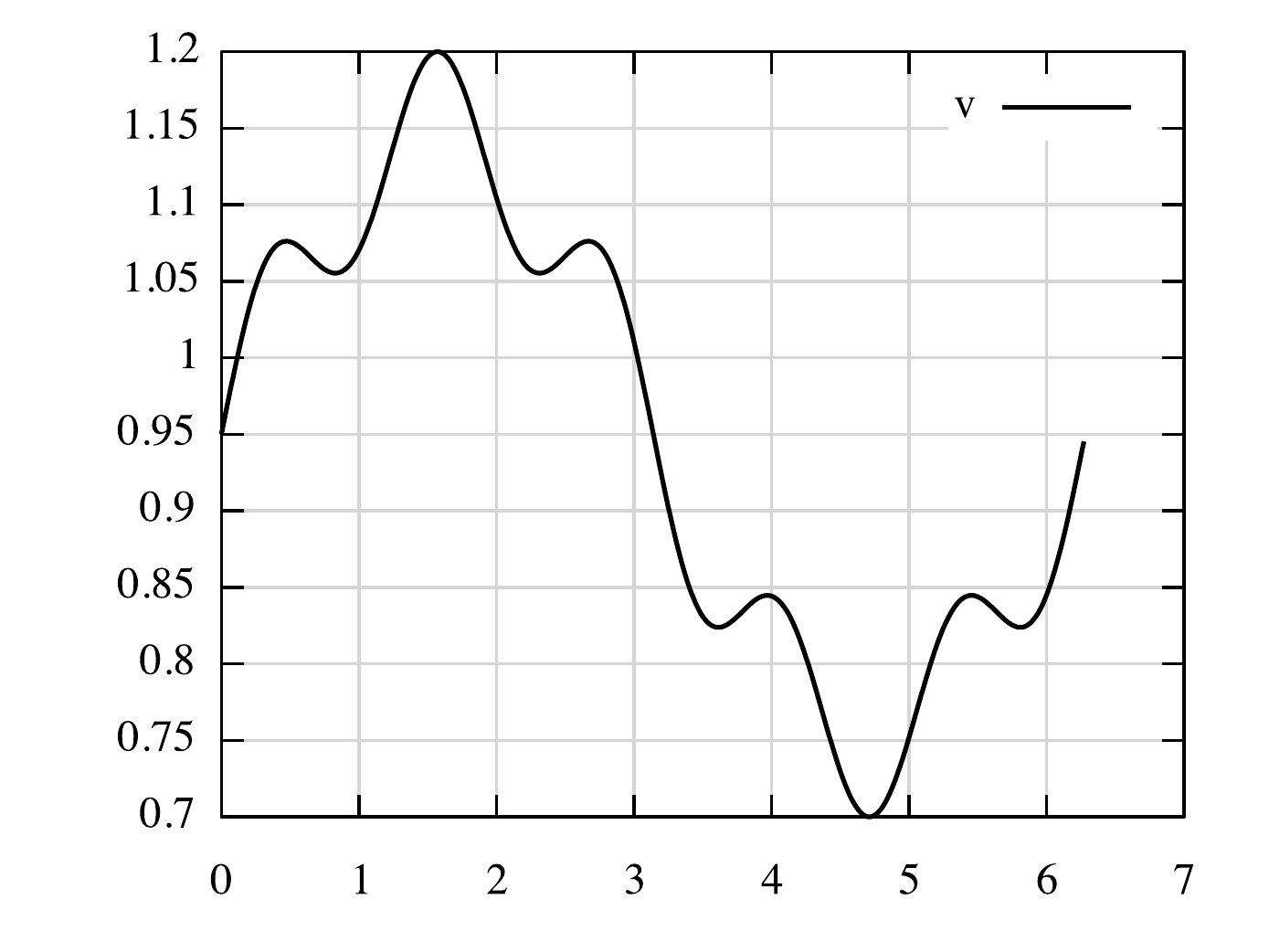}
\includegraphics[width=180pt,keepaspectratio=false]{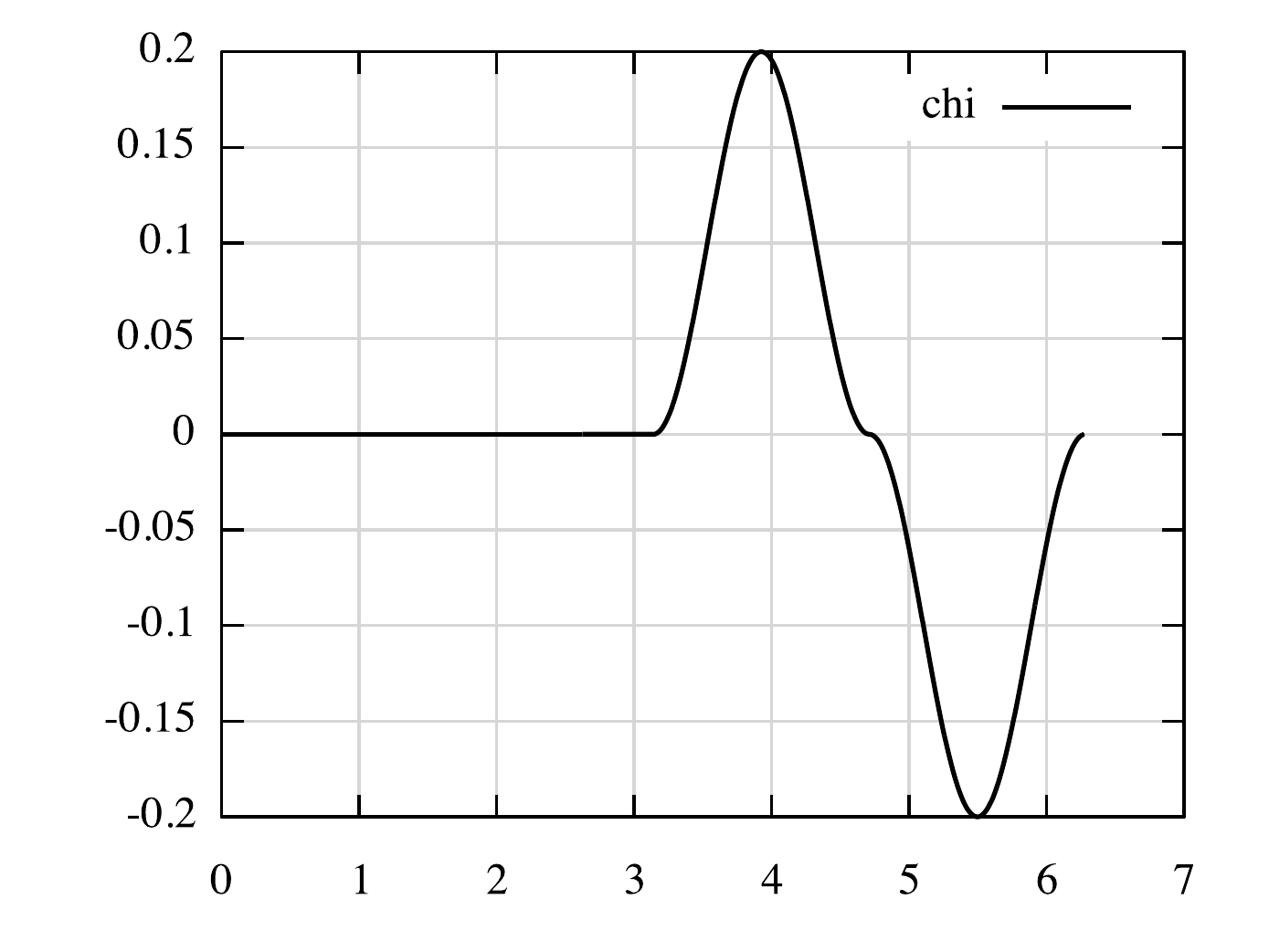}
$$
\caption{Initial data for the backward method with $v>1$ and $\chi=0$ on an interval.} 
\label{FIG.555}
\end{figure}

\begin{figure}[htbp]
$$
\includegraphics[width=180pt,keepaspectratio=false]{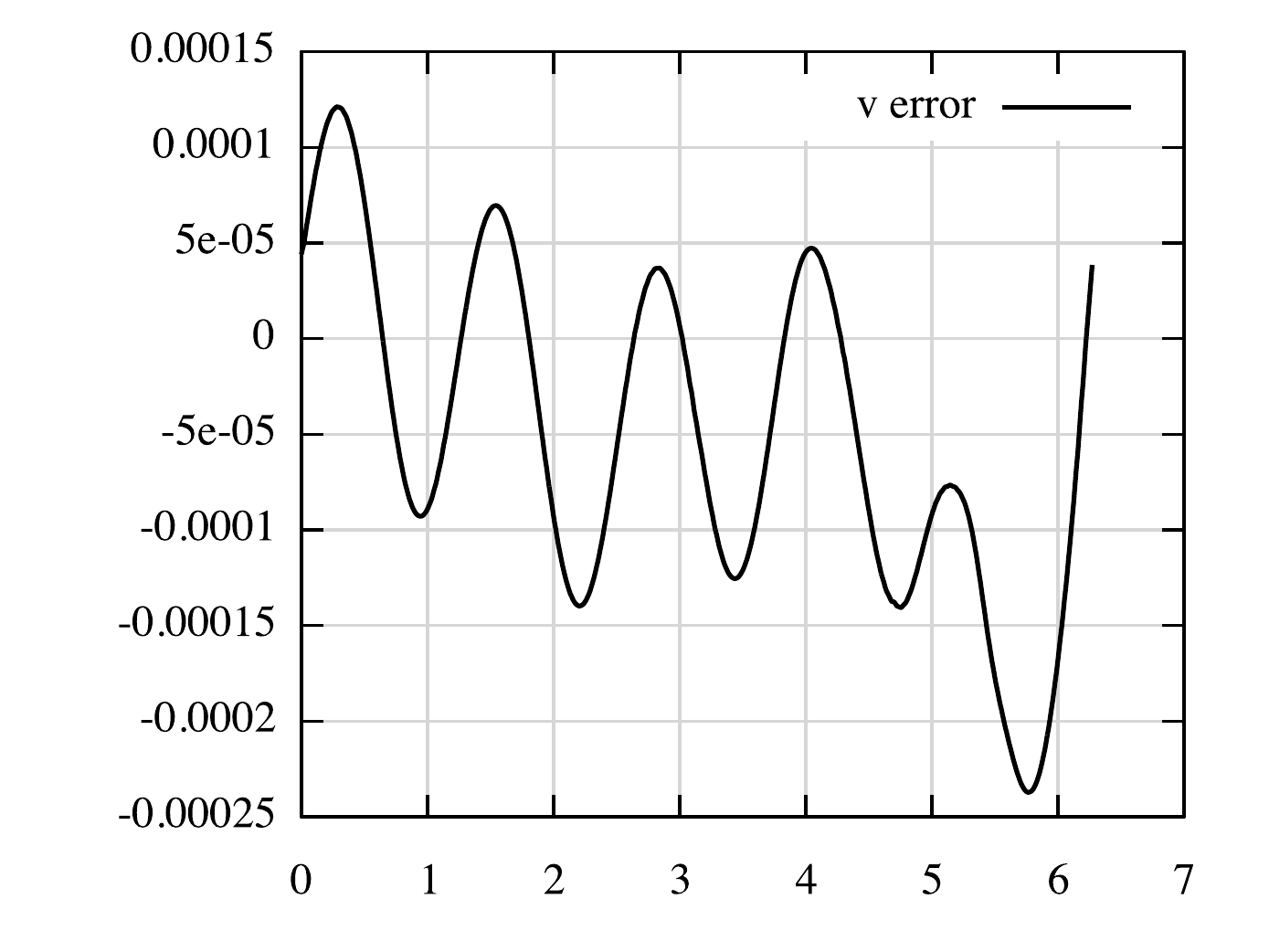}
\includegraphics[width=180pt,keepaspectratio=false]{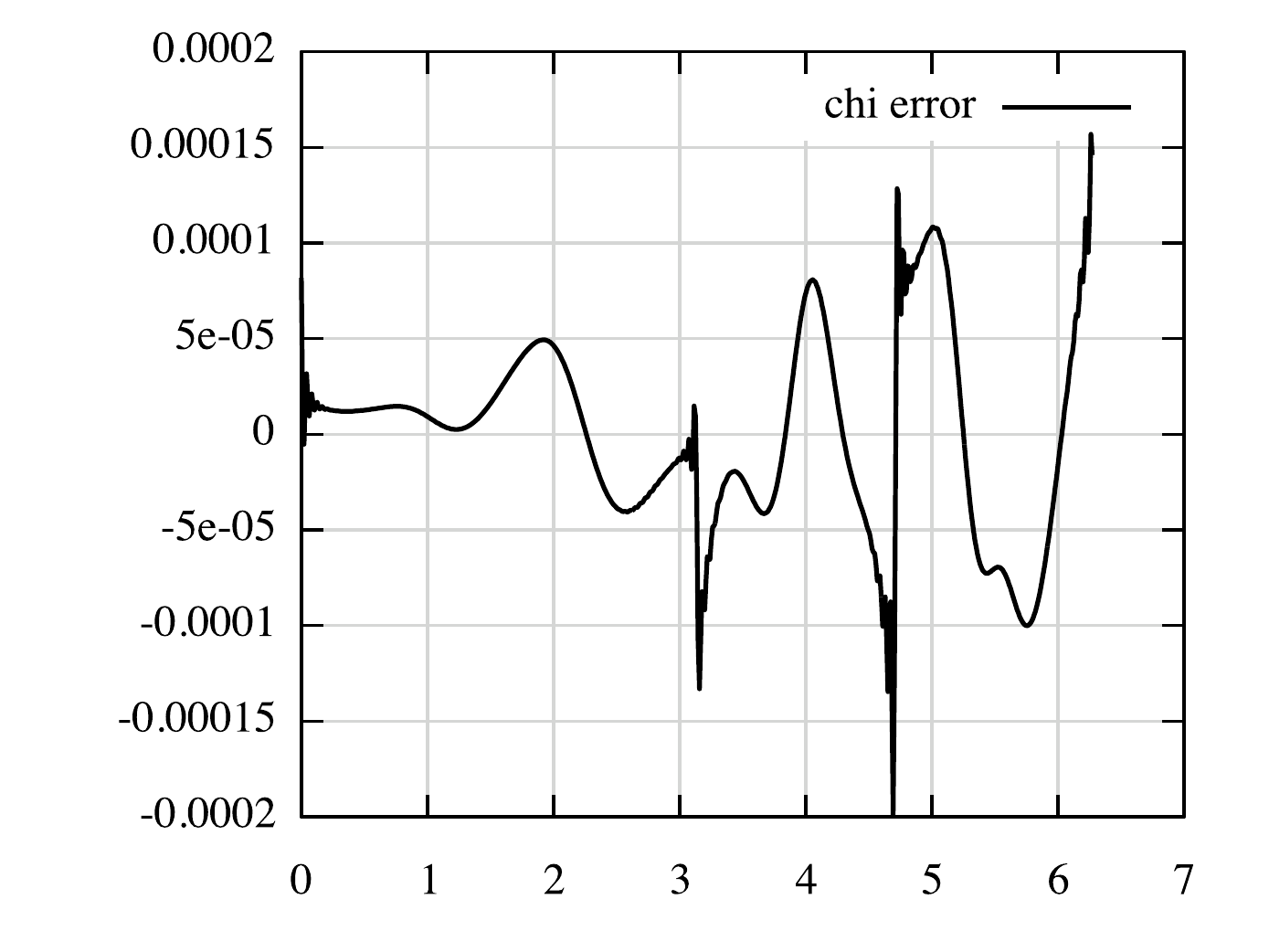}
$$
$$
\includegraphics[width=180pt,keepaspectratio=false]{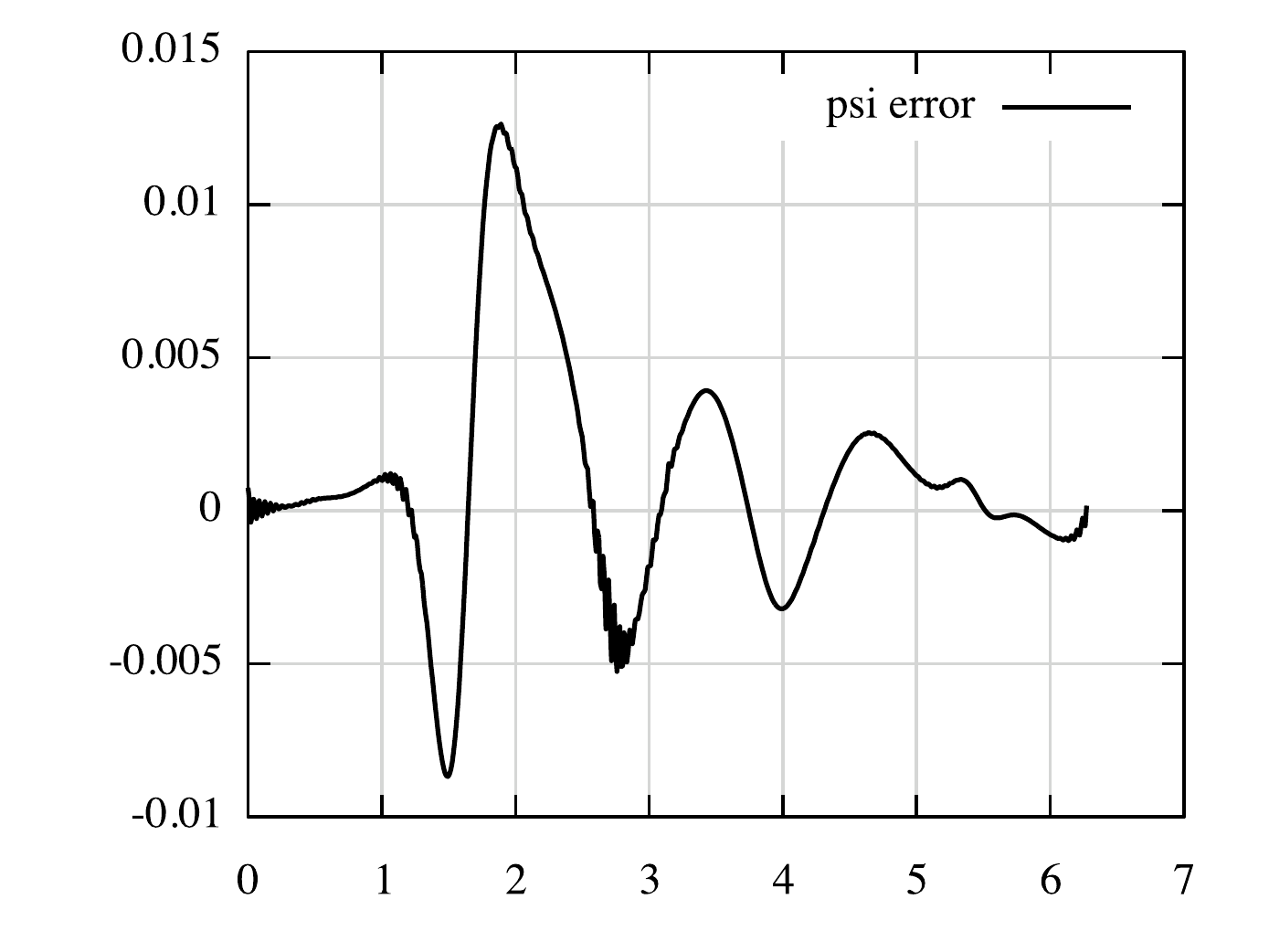}
\includegraphics[width=180pt,keepaspectratio=false]{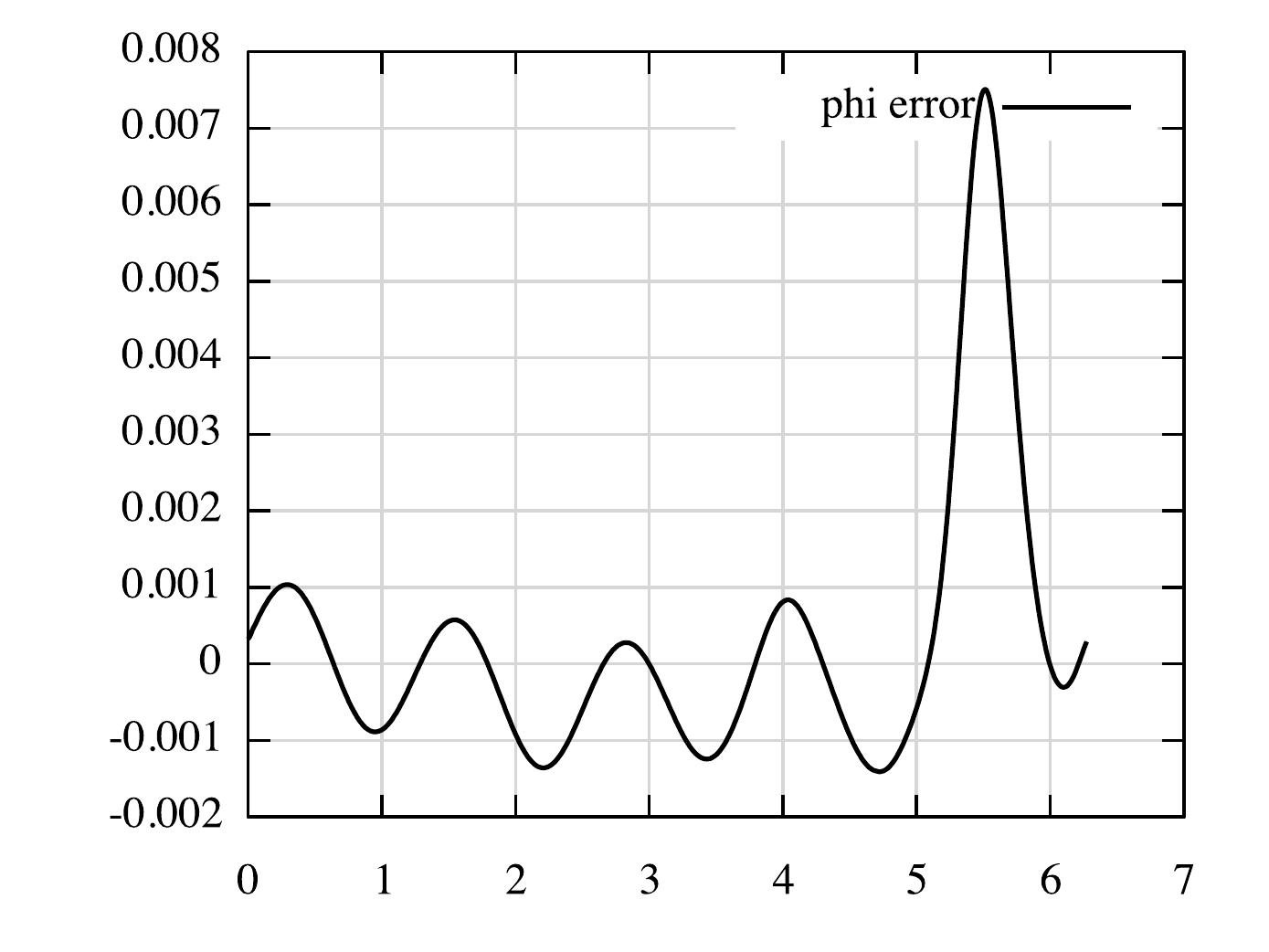}
$$
\caption{Error between $V,\Phi,\Psi,\chi$ given on the singularity and 
the numerical values after backward and forward evolutions, $\tau=11.5$, with $v>1$ and $\chi=0$ on an interval.}
\label{FIG.600}
\end{figure}

We can validate the heuristic analysis by relying on the proposed backward-forward strategy. 
We choose
the initial data 
$$
\aligned
&V(0) = 0.2\sin \theta+0.05 \sin 5\theta+0.95, 
\\
&  \Phi(0) = \cos \theta,
\\
&\Psi(0) = \sin (\theta-\pi/2) +0.76,
\\
&\chi(0) = \begin{cases}	
		0, & \theta\in(0,\pi),\\
		0.2 \mathop{\mathrm{sign}} (\sin 2\theta) \sin (2\theta)^2, &  \theta\in (\pi,2\pi), 
		\end{cases}
\endaligned
$$
where the velocity $v= V(0)$ is larger than $1$ on some interval; see Figure~\ref{FIG.555}. 
Accordingly, we set $\chi=0$ on that interval, but we 
note that $\chi$ does not vanish on the whole interval $[0,2\pi]$. 
The difference between these initial data and the result of the backward-forward simulation is plotted in Figure~\ref{FIG.600}. 

It is instructive to consider also a case where the backward-forward approach does
not accurately capture the behavior of the solutions. Namely,
by modifying the initial velocity $V(0)$ above, so that it is now larger than $1$ in some interval 
where $\chi$ does not vanish. According to our heuristics, it is not possible to obtain these data on the 
singularity by evolving towards it. Nevertheless, the backward method apparently does converge 
and generates some solution on the hypersurface $t=1$. By taking this numerical results as the initial data 
for the forward method and then evolving it toward the 
singularity, we have found that the computed solution ends up being very far from the initial data 
originally prescribed. This suggests that the solution computed by the backward code was physically meaningless. 
Our approach allows us to detect such a behavior.

%--------------------------------------------------------------------------------------------------------------------

\subsubsection{The case where $\chi, \Psi$ vanish at a single point}

The initial value $v$ for the backward code is now chosen to lie outside the interval $(0,1)$ at one point, only. 
The purpose here is to show that our method can accurately simulate these exceptional situations.
We perform the backward-forward simulation and compare the result to the prescribed velocity. 

\begin{figure}[htbp]
$$
\includegraphics[width=180pt,keepaspectratio=false]{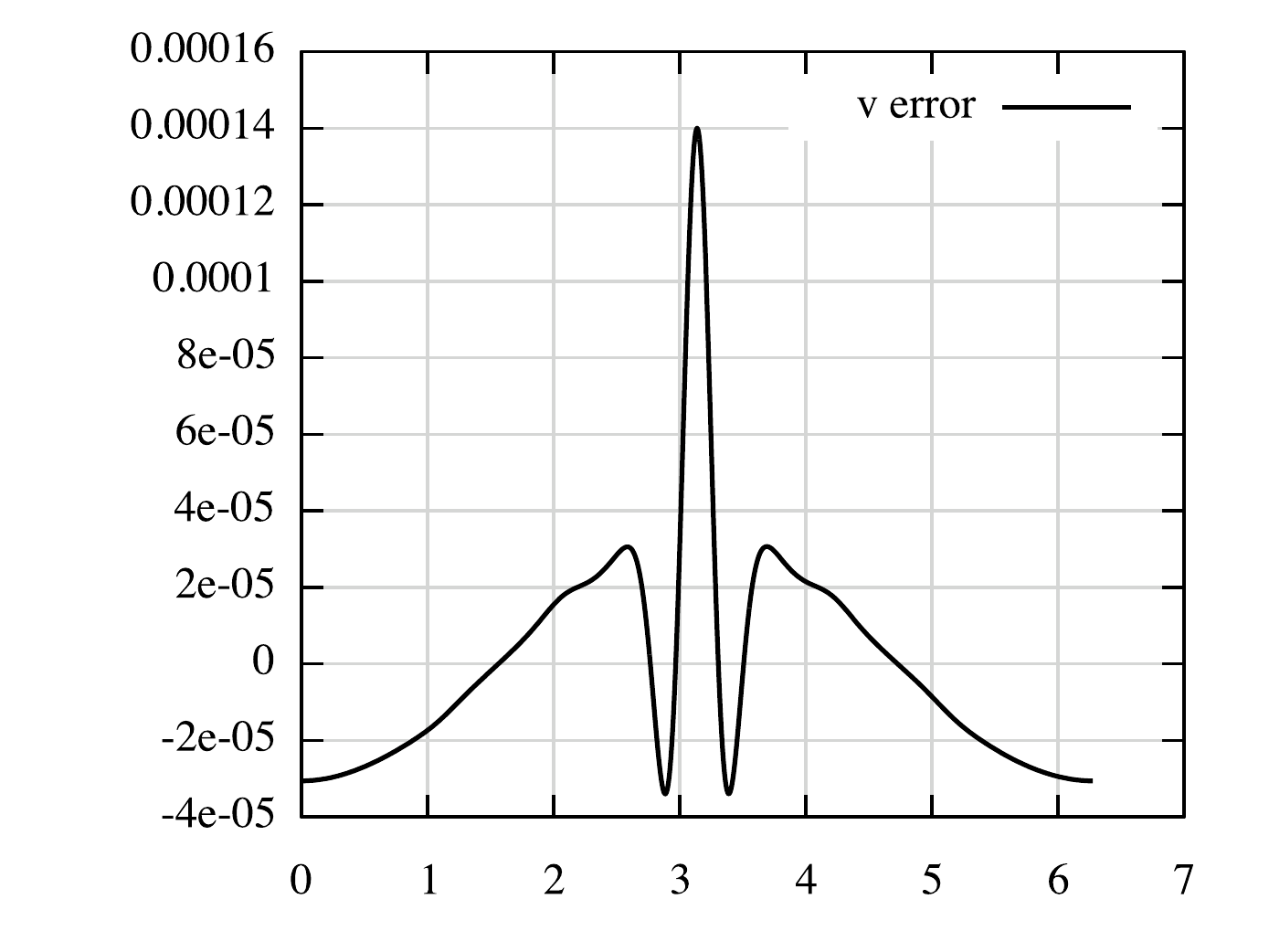}
\includegraphics[width=180pt,keepaspectratio=false]{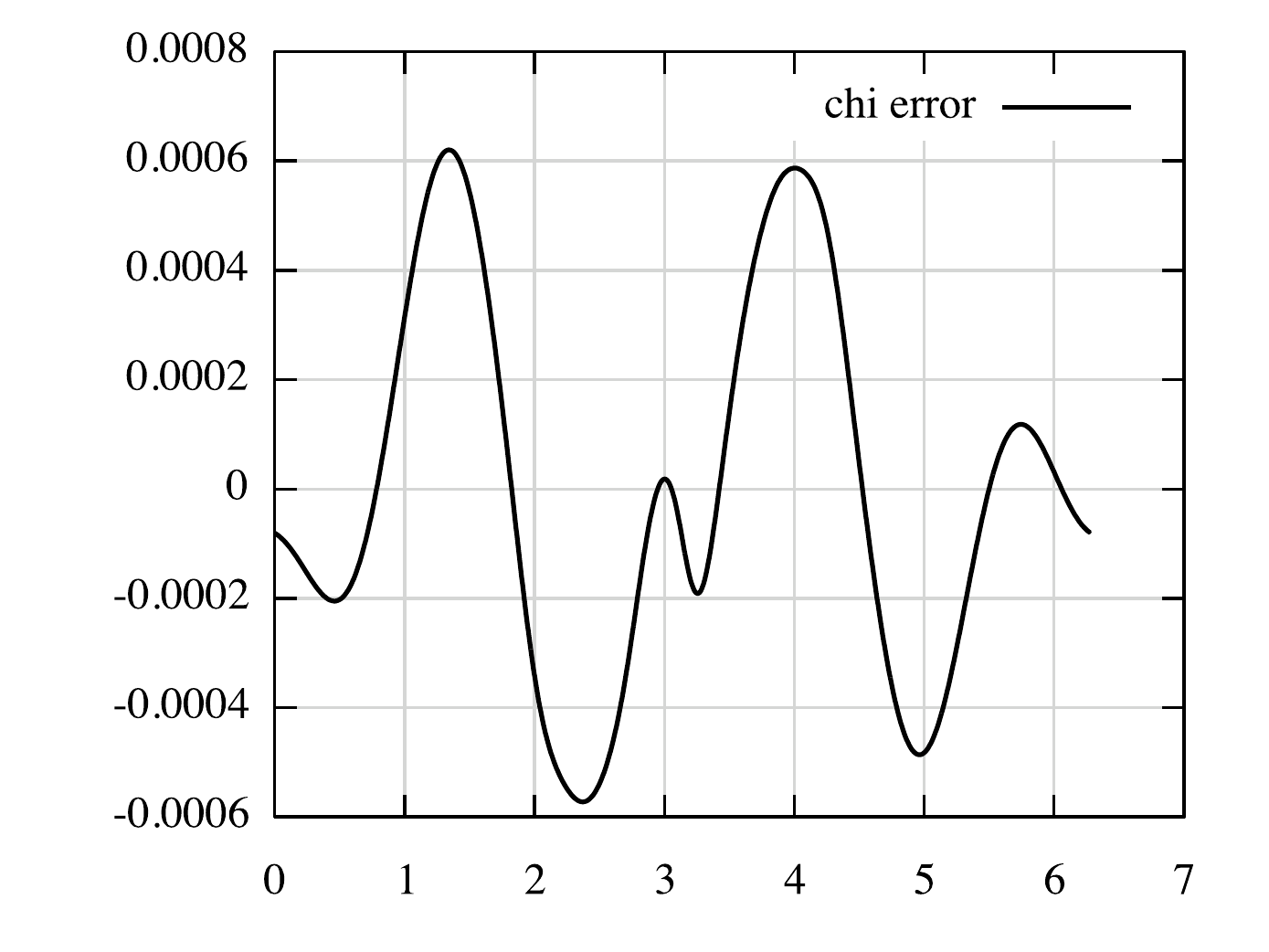}
$$
$$
\includegraphics[width=180pt,keepaspectratio=false]{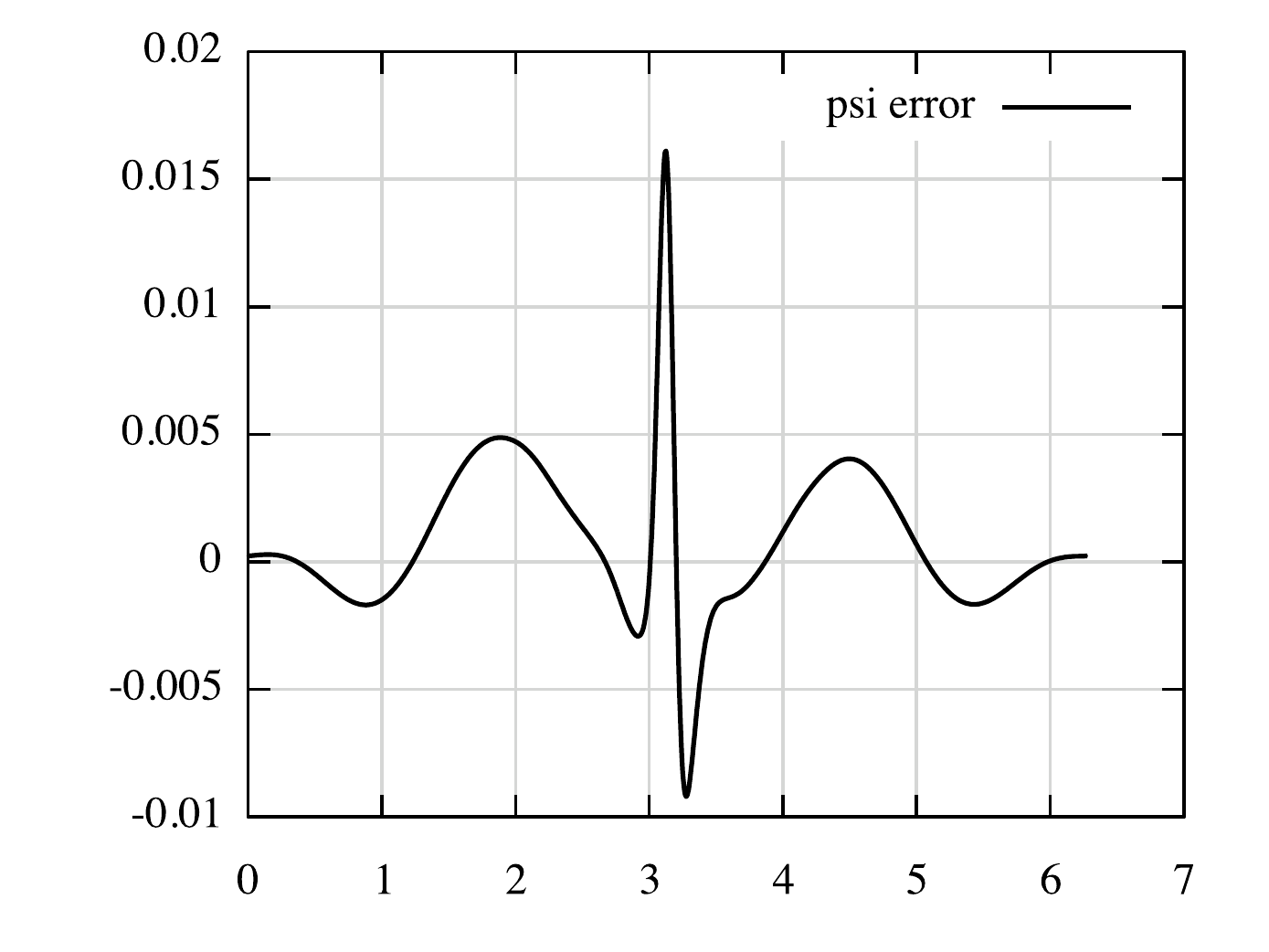}
\includegraphics[width=180pt,keepaspectratio=false]{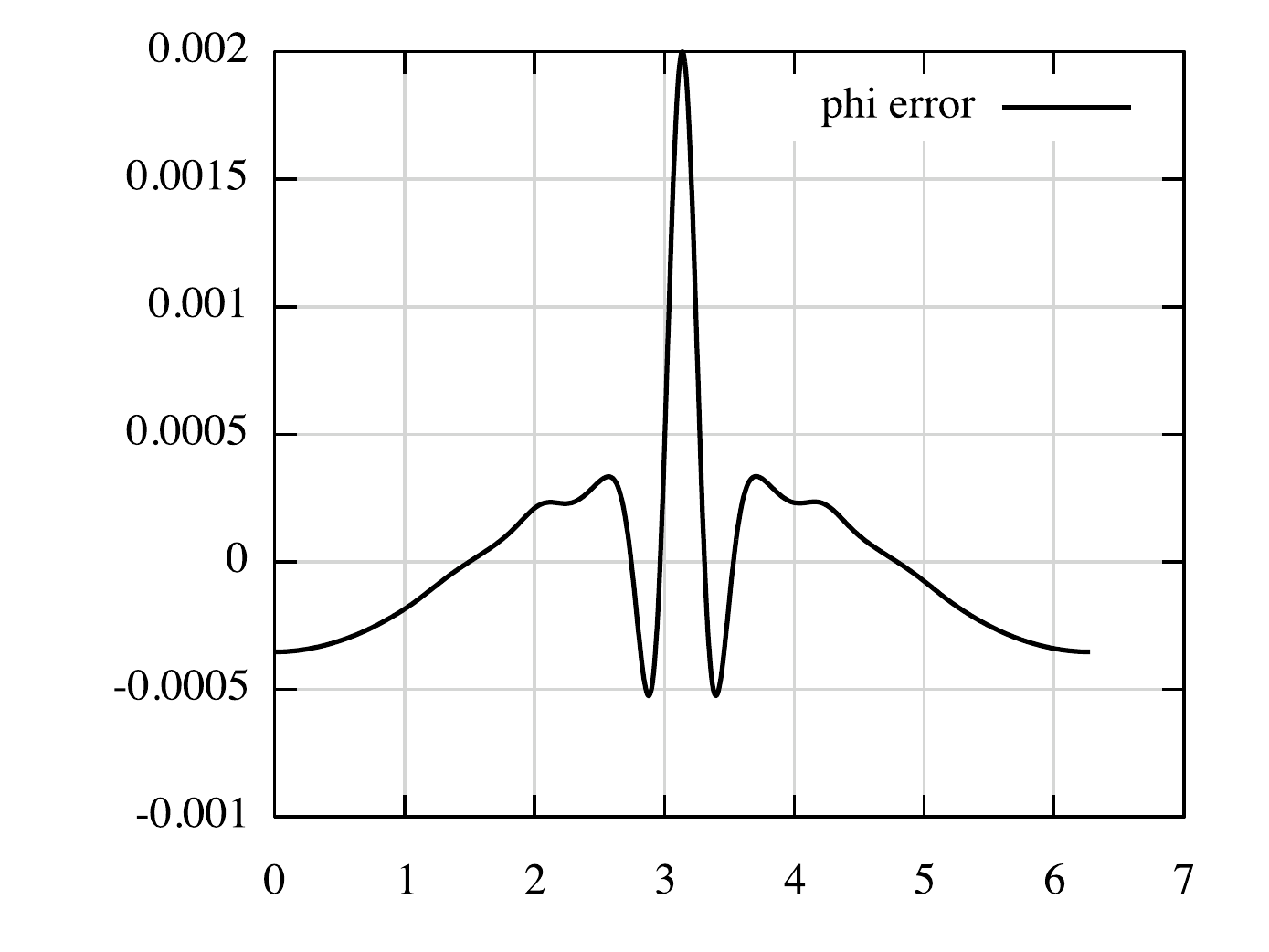}
$$
\caption{Error between $V,\Phi,\Psi,\chi$ given on the 
singularity, and numerical values after backward and forward evolutions, $\tau=11.5$, for $v =1$ at one point.}
\label{FIG.690}
\end{figure}

The initial data are taken to be 
\be
\label{BCP.45}
\aligned
&V(0;\lambda) = \frac{1}{2} e^{-\lambda(\theta-\pi)^2} + 0.5, \qquad &&\Phi(0) = \cos \theta +C,
\\
&\Psi(0) = \sin 2\theta, \qquad &&\chi(0) = \frac{1}{2} (1- \cos 2\theta)^2 \mathrm{sign} (\pi-x),
\endaligned
\ee
where $\lambda=20$, and the constant $C$ is chosen so that the constraint \eqref{GS.Einstein3} holds. 
At the point $\theta=\pi$ 
where $v$ equals $1$, both $\chi$ and $\chi_\theta$ vanish, as required by the heuristics. 
The error between the computed quantities after the backward-forward simulation and the initial data for 
the backward method is shown in Figure~\ref{FIG.690}.

This initial data chosen for the velocity $v$ is intended to qualitatively resemble the spiky features observed 
in Gowdy spacetimes, which are discussed below in more detail.

It is interesting to also consider various values of the parameter $\lambda$, and to look for a possible limit 
beyond which our simulations are no longer accurate. As $\lambda$ grows, the asymptotic velocity $V(0;\lambda)$ 
approaches the constant $1/2$, except at $\theta=\pi$, where it takes the value $1$. 
This is thought to be the asymptotic behavior of the spikes on the velocity.

It is natural that our method breaks down beyond a certain value of $\lambda$, since the initial velocity 
is then approaching a {\sl singular function.} Still, we have found that the approximation by the backward-forward method 
still works up until about $\lambda=900$ (with $K=500$ and time step $0.0005$). However, this accuracy is improved by taking 
larger values of $K$ and smaller values of the time increment.

\begin{figure}[htbp]
$$
\includegraphics[width=180pt,keepaspectratio=false]{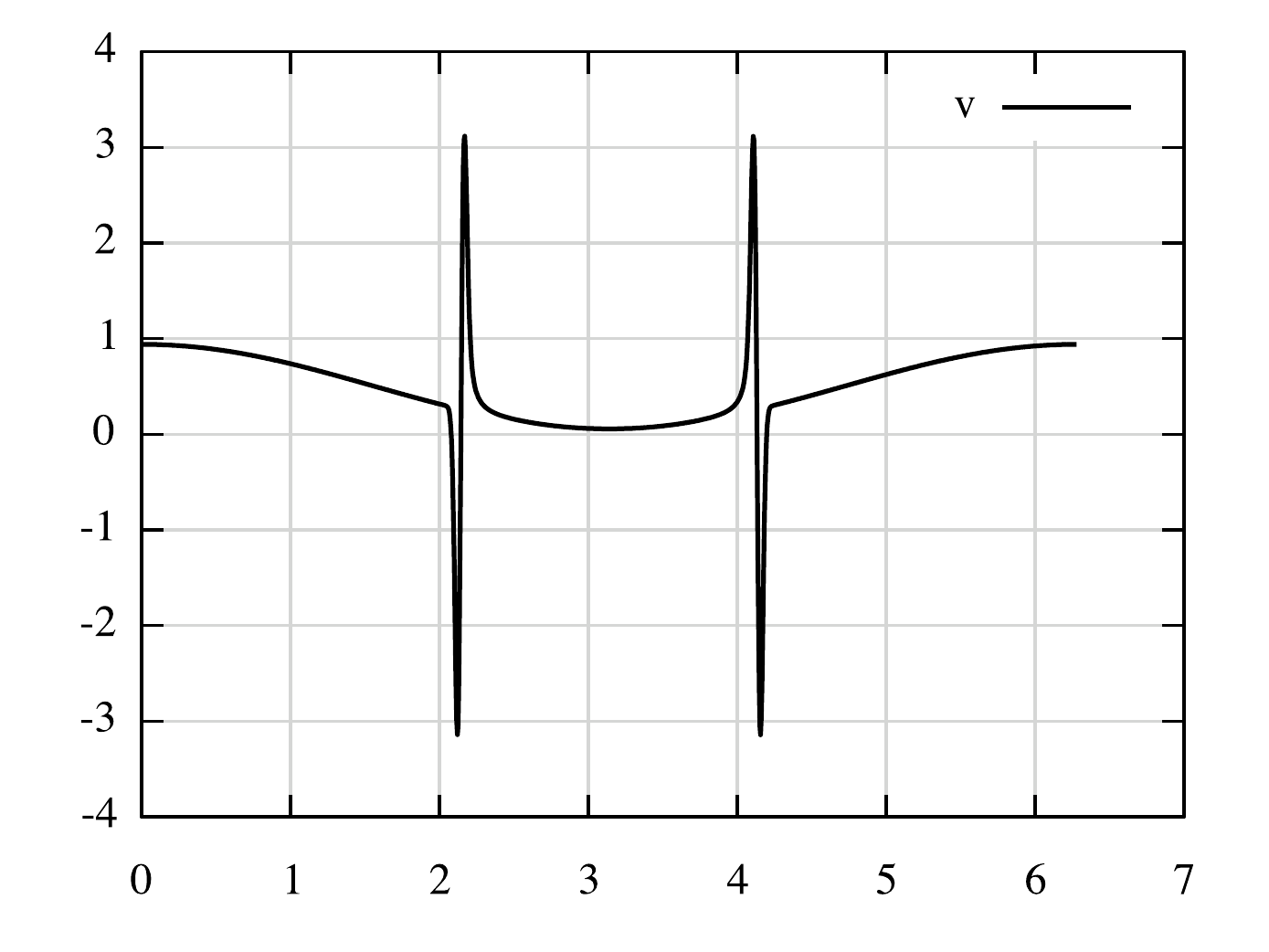}
\includegraphics[width=180pt,keepaspectratio=false]{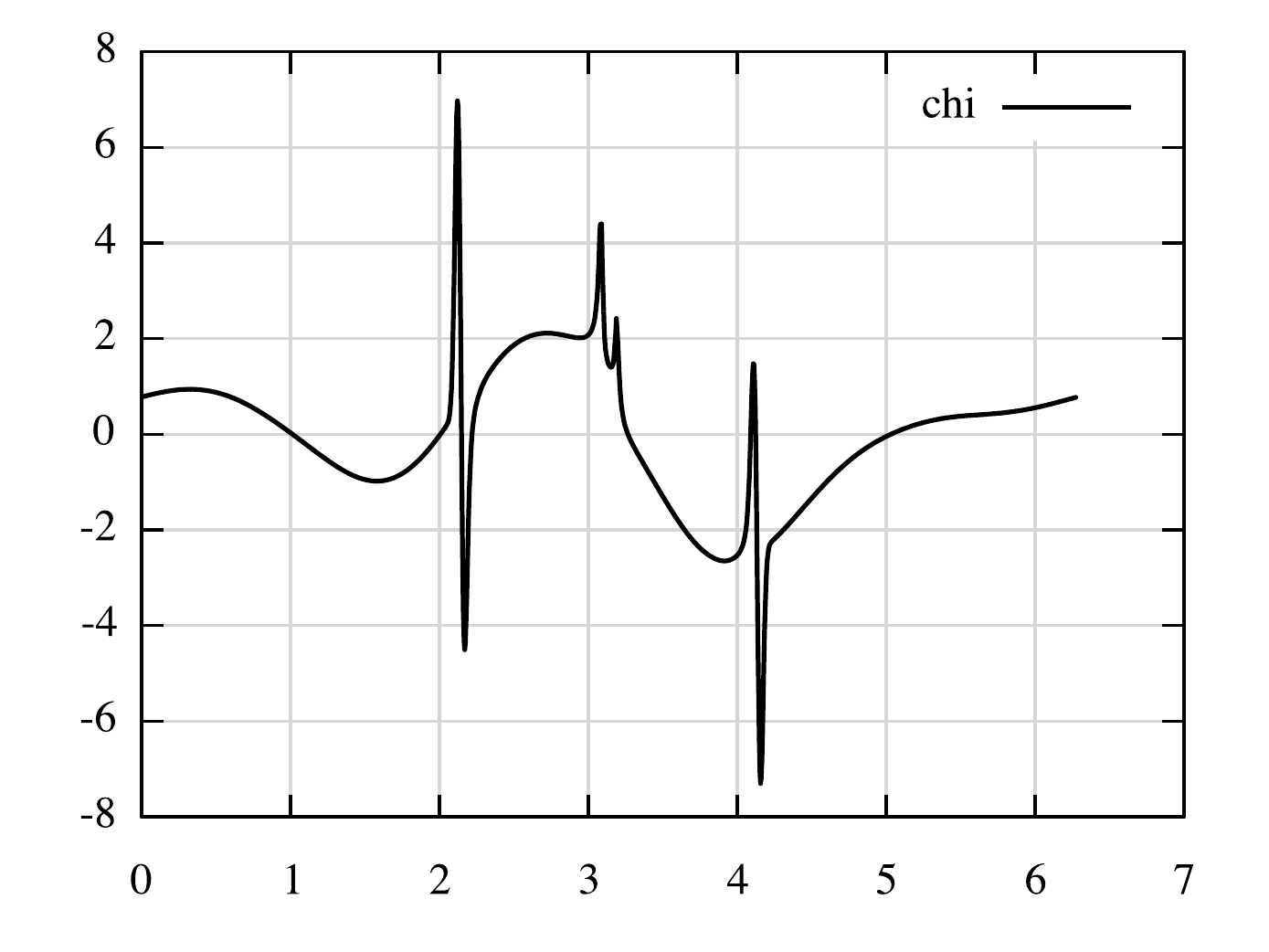}
$$
$$
\includegraphics[width=180pt,keepaspectratio=false]{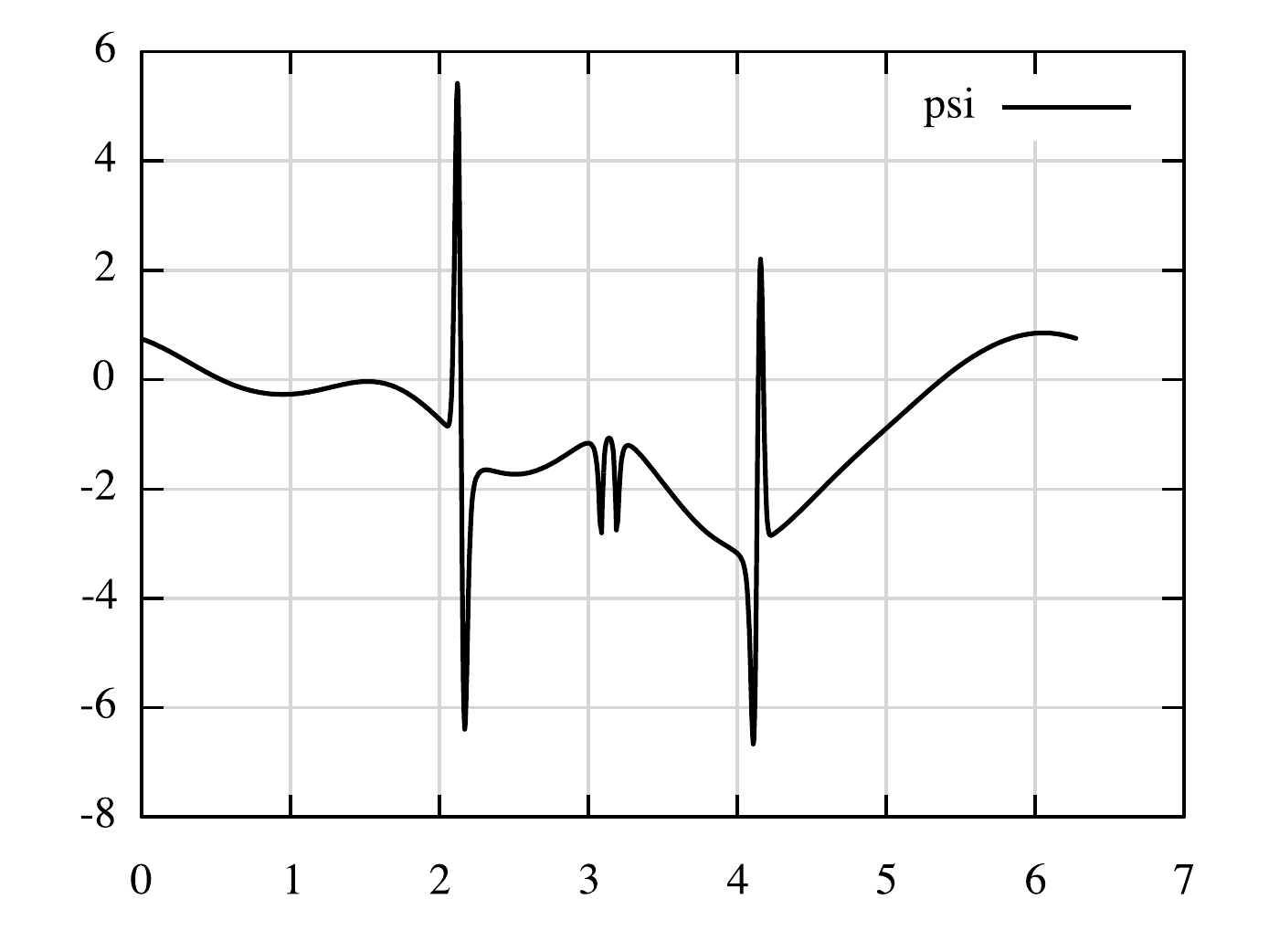}
\includegraphics[width=180pt,keepaspectratio=false]{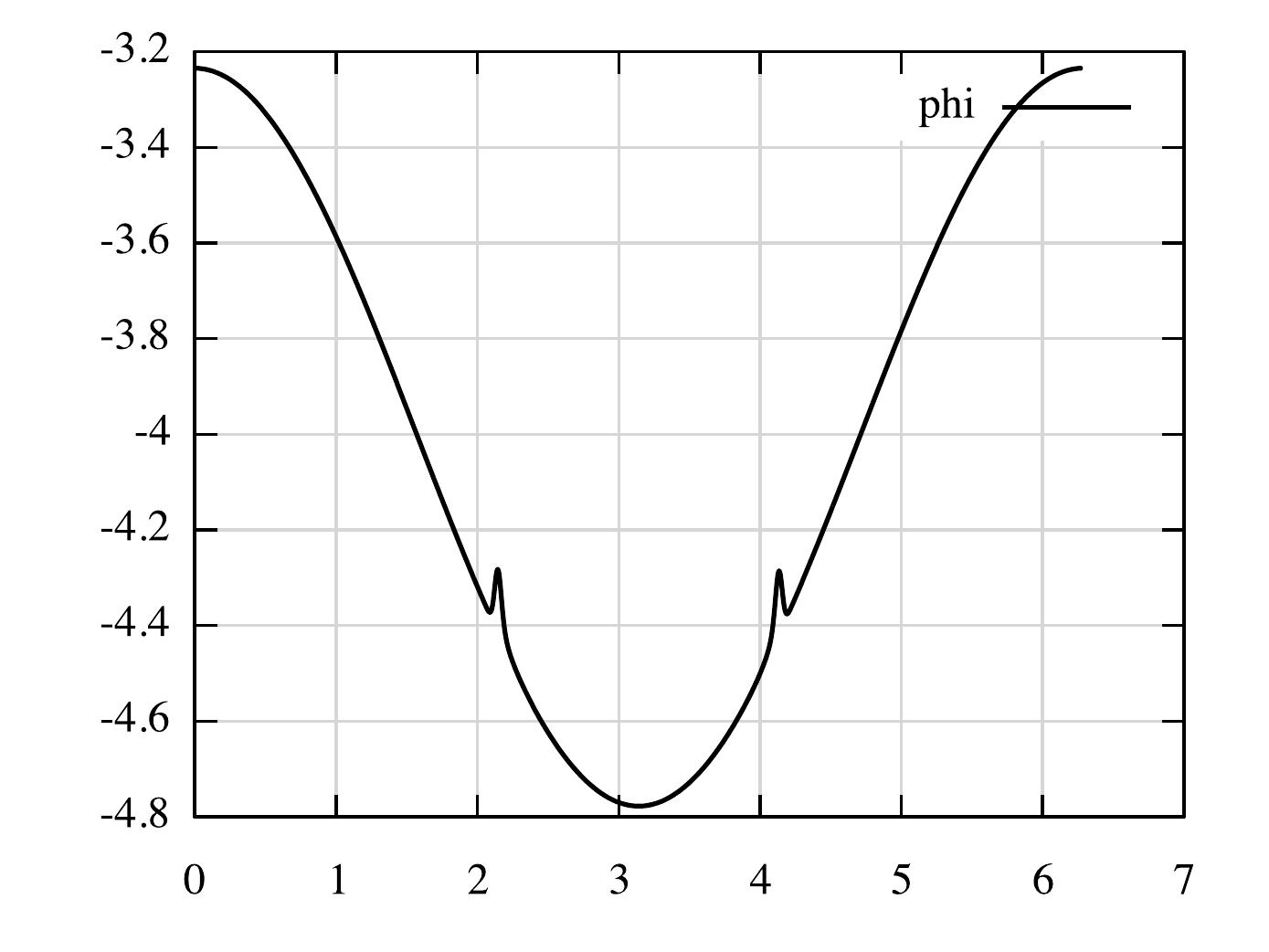}
$$
\caption{Solution computed by the backward code at $t=1$ ($\tau=0$) with initial data given by \eqref{BCP.45}.
 (This solution is taken as the initial value for the forward code.)}
\label{FIG.200}
\end{figure}

\begin{figure}[htbp]
\begin{center}
$$
\includegraphics[width=180pt,keepaspectratio=false]{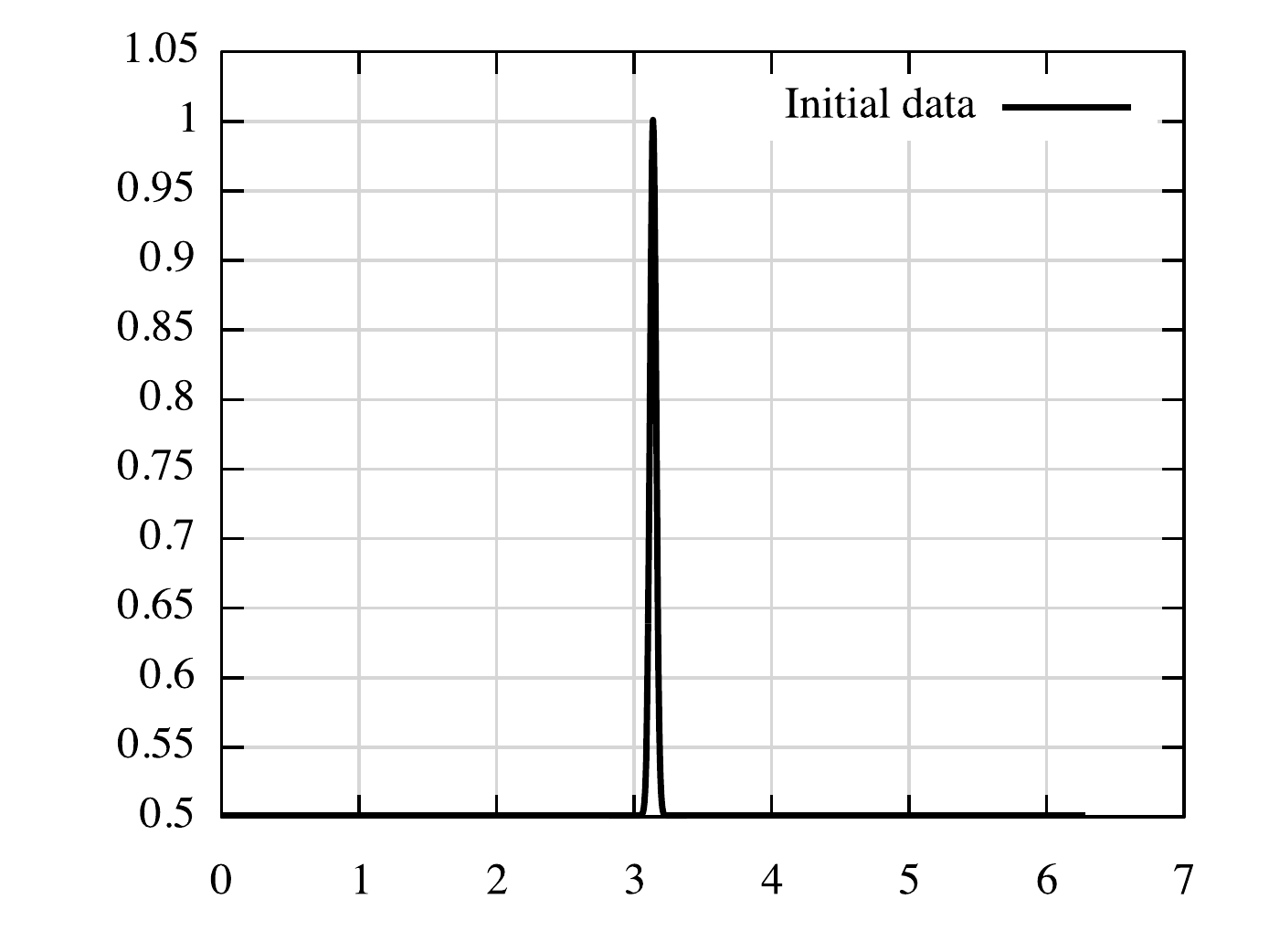}
\includegraphics[width=180pt,keepaspectratio=false]{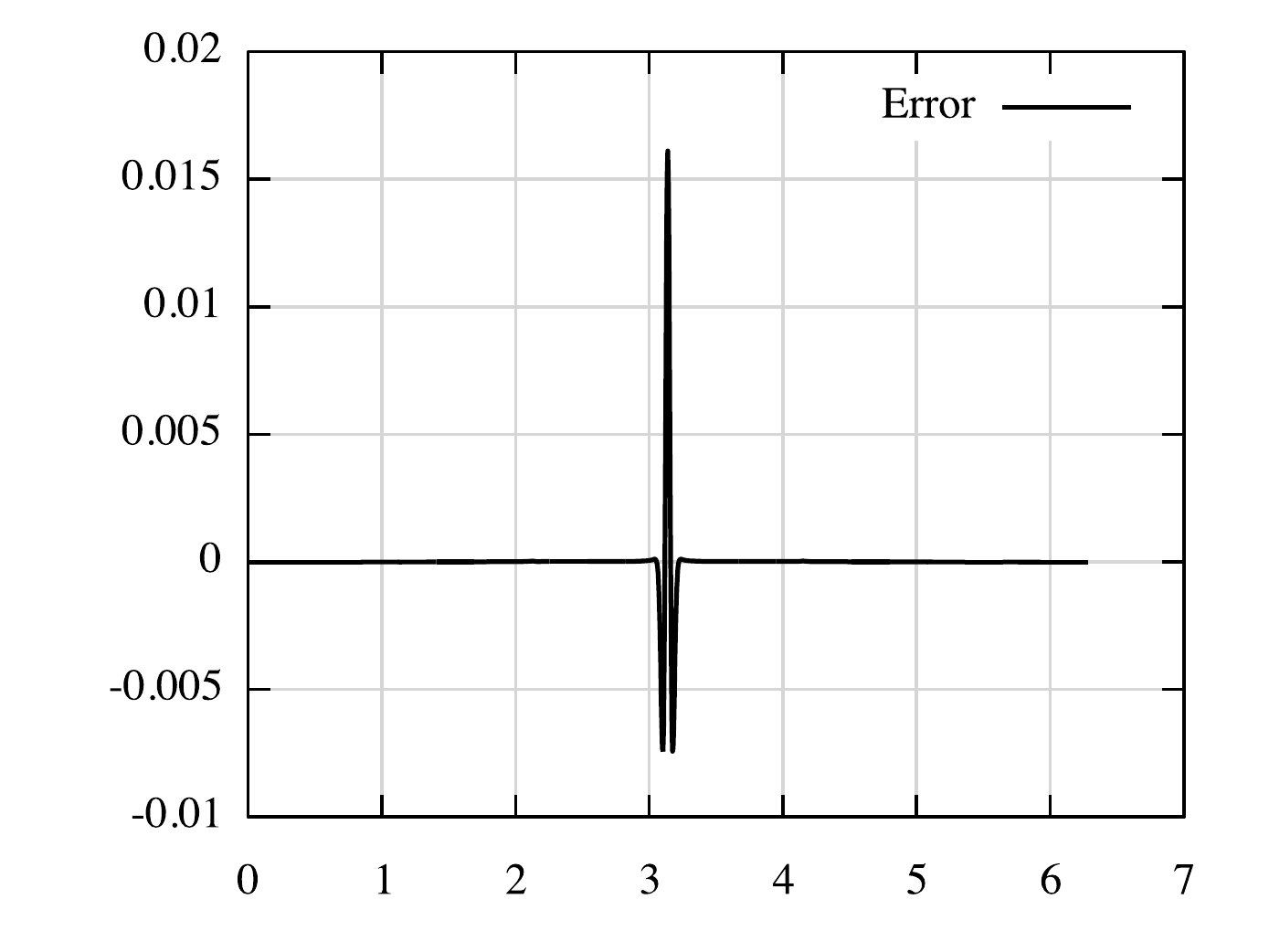}
$$
\caption{Spiky initial asymptotic velocity, $t=0$ (left), and the error 
between the computed asymptotic velocity after backward and forward evolutions and the initial data, $\tau=10$.}
\label{FIG.250}
\end{center}
\end{figure}

The solution of the backward simulation at $t=1$ for the initial data \eqref{BCP.45} and $\lambda=1000$ is shown 
in Figure~\ref{FIG.200}. The sharp gradient features observed in the solution at the time $t=1$ 
are simply 
due to the fact that the data itself 
was chosen to be have a sharp gradient on the singularity. 
In Figure \ref{FIG.250}, we show the corresponding spiky upward-pointing initial data $v$ (left) 
and the error between the numerical velocity after a backward-forward simulation and the initial data for the backward method (right).
In this case, the error in the variables $\Psi, \chi$ is rather large, showing that, in these more challenging cases,
some information is lost in the backward-forward evolution. Nevertheless, this discrepancy between the initial data for $\Psi, \chi$ 
and the values computed after the backward-forward evolution is {\sl localized on small regions}
of the interval $[0,2\pi]$. The functions $\Psi(0)$ and $\chi(0)$ are otherwise well approximated. 

Since our main goal, at this stage,
 is not to provide a competitive numerical scheme for Gowdy spacetimes, but rather to demonstrate
the validity of the proposed strategy, we argue that  
this discrepancy does not invalidate our strategy. It merely shows that our method breaks down when dealing with steep, 
singular data --which is not surprising from the standpoint of spectral approximation theory. 

The situation here is quite different from the one described in the previous paragraph, 
where we \emph{expected} a large difference between, say, the data $\Psi(0)$ and the result 
of the backward-forward evolution. 
In that case, we observed that the two resulting functions were completely unrelated. 
In contrast, in the present situation our approach preserves the overall shape of the functions, at least,  
and is also a rather good approximation in most regions.

Finally, note that the asymptotic velocity $v$ (which is well approximated) is the most important quantity from the physical standpoint, 
since (see \cite{KR}) $v$ is in fact the only coefficient which determines the nature of the coordinate 
singularity: the value of $v$ alone determines whether the curvature blows up as one approaches the
coordinate singularity. Therefore, even in these cases where our method does not provide an accurate approximation 
of certain coefficients, the physically relevant coefficient $v$ is always well approximated, at least in the variety of cases we tested.

\begin{figure}[htbp]
$$
\includegraphics[width=180pt,keepaspectratio=false]{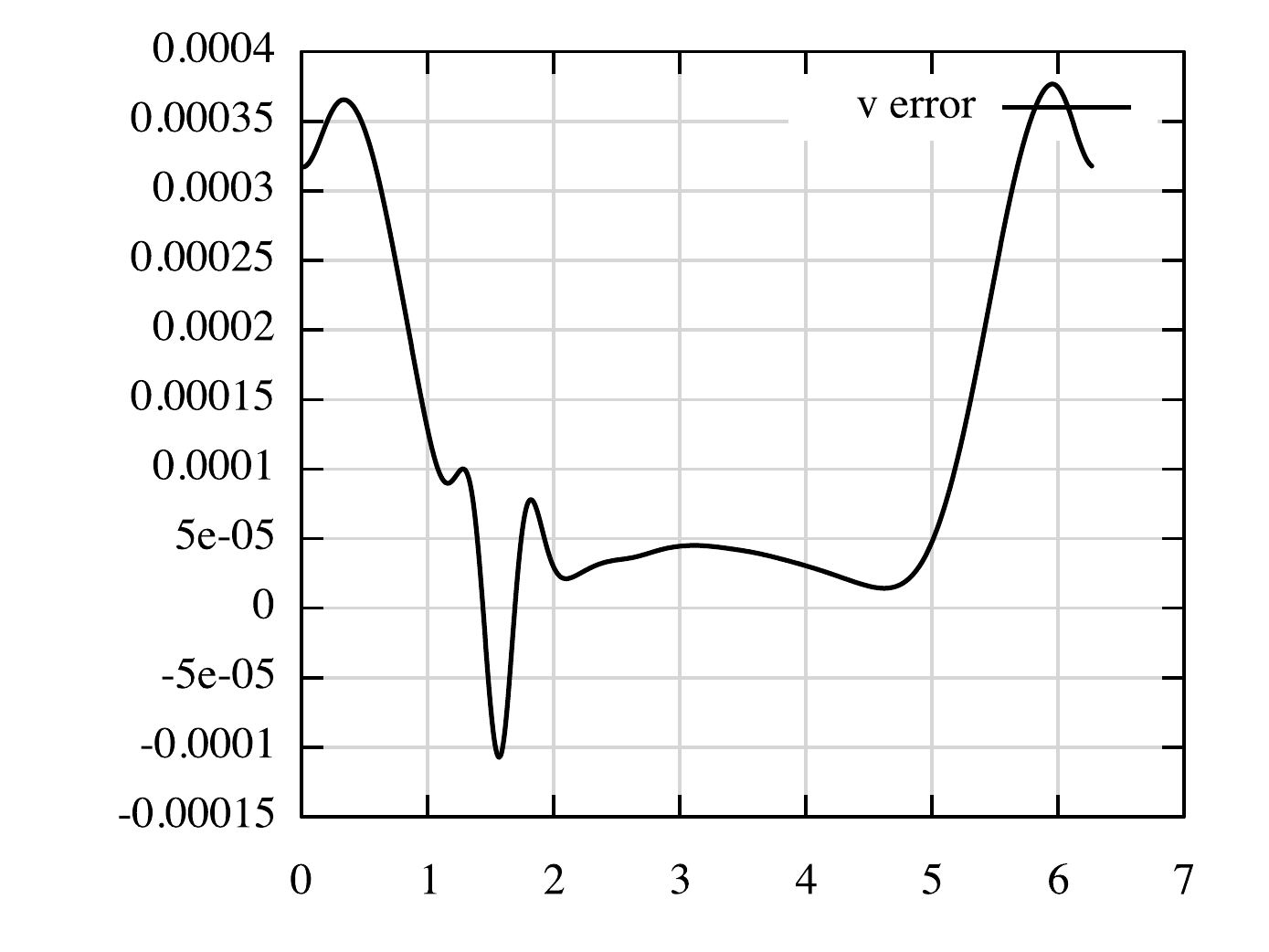}
\includegraphics[width=180pt,keepaspectratio=false]{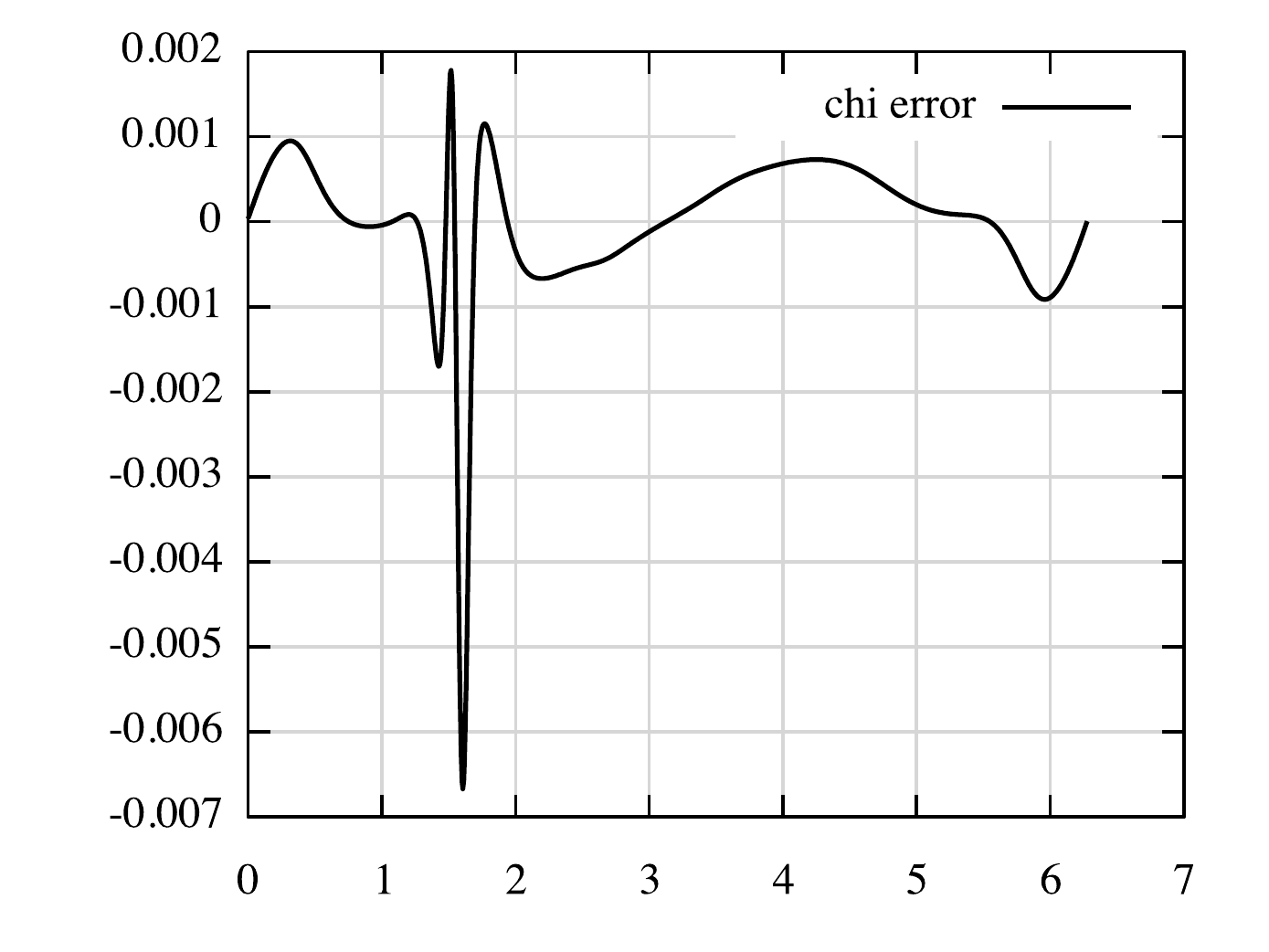}
$$
$$
\includegraphics[width=180pt,keepaspectratio=false]{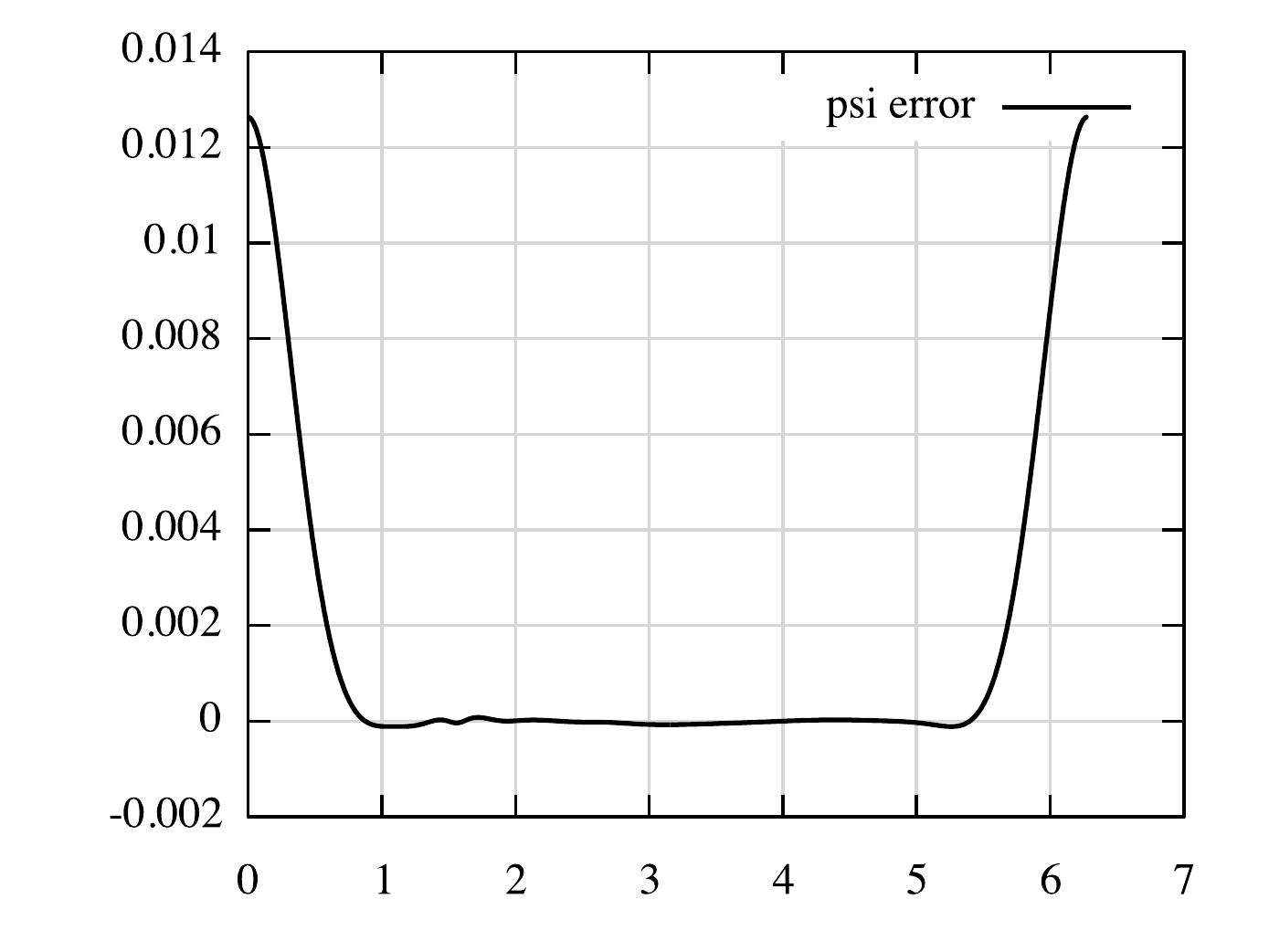}
\includegraphics[width=180pt,keepaspectratio=false]{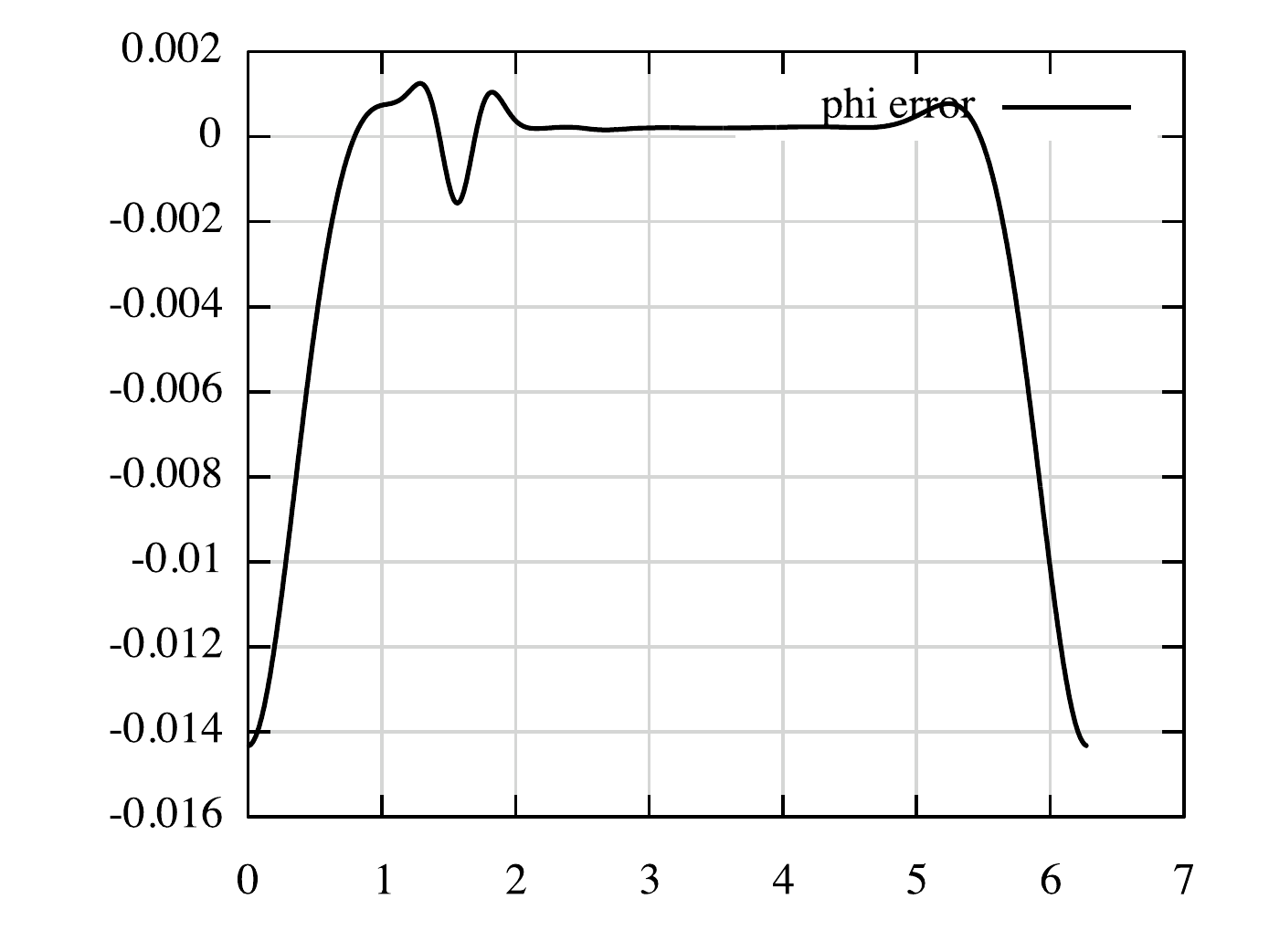}
$$
\caption{Error between $V,\Phi,\Psi,\chi$ given on the 
singularity, and numerical values after backward and forward evolutions, $\tau=11.5$, for $v =0$ at one point.}
\label{FIG.700}
\end{figure}

The results obtained with initial data satisfying $v \leq 0$ at some point 
are similar to the ones for $v \geq 1$ but, now, the data $\Psi(0)$ must be such that both $\Psi, \Psi_\theta$ vanish at those points.
The error between the initial data and the computed backward-forward quantities is plotted in Figure~\ref{FIG.700}, 
for initial data similar to \eqref{BCP.45} and with $\lambda=20$.

%=========================================================================================================================================================

\subsubsection{Relation to spikes on the velocity}

Since the points where $\Psi$ or $\chi$ vanish are, in general, isolated points, 
the asymptotic velocity $v$ generated by the forward code 
(and, in turn, the metric variable $P$)
 may exhibit \emph{spikes} at 
these exceptional points. See, for instance, Figure~\ref{FIG.450}. 
This happens because the solution $V$ of the forward problem \eqref{SN.5} (which tends to $v$ as $t \to 0$) is forced 
to enter the interval $(0,1)$ {\sl everywhere except} at these isolated points, where it may remain outside the interval. 
This is the origin of the spiky features observed in Gowdy spacetimes. These spikes have been extensively studied both numerically 
and theoretically. 

The main purpose of the previous simulations was to demonstrate that the forward-backward strategy allows one 
to deal with situations involving spikes, while still maintaining an acceptable level of accuracy. 
We emphasize again that the fact that accuracy breaks down when the spiky initial formation is sufficiently steep
is completely natural, since our method needs sufficiently smooth data to work well, 
and that given sufficient spatial resolution and computing power, we should be able to accurately simulate any 
given spiky feature by the forward-backward strategy. 

%=========================================================================================================================================================

\subsubsection{Limitations and extensions of the method}

As a natural extension of our approach we could choose some initial data for the forward method, such as \eqref{BCP.42}, 
then evolve it to a large time $\tau$, and finally apply the backward method to the result. 
As established in the present paper, this strategy is successful
for initial data originating from solutions without spikes. However, our formulation does not allow us to consider 
more general data. 
To see this, observe that, in Figure~\ref{FIG.450}, the metric coefficient $Q$ appears to develop a discontinuity near $\theta=1.5$. 
This means that $Q_\theta$ will blow up on the coordinate singularity. On the other hand, in view of the definition of the unknowns for the backward method 
(see \eqref{BCP.8}),  this means that $\chi$ does not take a finite value on the 
singularity. Thus, our backward method does not allow us to simulate such solutions.

In principle, it should be possible to use the forward-backward strategy and design a numerical method covering such cases. 
We could define new unknowns (in the spirit of \eqref{BCP.8} but) taking into account higher-order terms in the corresponding expansions. 
For instance, we could introduce 
$$
\overline \chi : = Q_\theta - \big( t^{2v} \psi \big)_\theta,
$$
which, according to the asymptotic expansions \eqref{GS.expa}, should now converge to a finite limit,
 even in cases such as the one in Figure~\ref{FIG.450}.

%=======================================================================================================================

\section{Concluding remarks}

The numerical simulation of (coordinate) singularities in general relativity is a particularly 
challenging problem, from both the mathematical and the computational standpoints. 
In the present paper, we have initiated a new strategy for the numerical investigation of $T^3$ Gowdy spacetimes:  

\begin{itemize}

\item We have recovered classical results using a Fourier pseudo-spectral scheme which produces accurate results 
using rather modest computing power. Indeed, these simulations were performed on a MacBook Pro at 2.4GHz, using the free software Scilab.
The running time was of the order of 2--3 hours for the longest computations. 

\item We have designed a new backward scheme, also based on pseudo-spectral approximation,  
which is computationally undemanding and allows to simulate the asymptotic behavior by considering an initial-value problem 
with ``final data'' on the coordinate singularity. 

\item This backward approach was validated by first evolving away from the 
singularity (using the backward code) and then evolving back toward the 
singularity (using the forward code). 

\end{itemize}

After a backward-forward evolution, we have obtained accurate results, 
and we firmly established the backward approach as a reliable approach to study Gowdy spacetimes.  
Interestingly enough, the proposed backward-forward strategy fails to work in the situations where 
it is not assumed to work, that is, when we set as ``final data'' on the 
singularity some data which are not allowed by the heuristics described in Section~\ref{HR}.

The Einstein equations for Gowdy spacetimes are known to produce solutions with non-smooth behavior on the
coordinate singularity (cf.~the spikes discussed in the previous section).
Through extensive numerical experiments 
we have established 
that the backward-forward simulation is valid even 
with initial (backward) data which are ``almost spiky'', while still retaining a good quality of the results. 
Our method allows us to cover a large class of spikes, and, moreover,
suggestions have been made on how to extend the proposed scheme 
and possibly deal with even more singular regimes. 

In short, the proposed backward strategy leads to a robust numerical method which 
allows us to accurately simulate the behavior of a large class of Gowdy spacetimes near the singularity.

%==========================================================================================================

\section*{Acknowledgements}

The third author (PLF) is thankful to the organizers (P.T. Chrusciel, H. Frie\-drichs, P. Tod) 
for their invitation to participate to 
the Semester Program ``Global Problems in Mathematical Relativity'' which took place at the Isaac 
Newton Institute of Mathematical Sciences (Cambridge, UK) in the Fall 2005 
and where this research was initiated. This paper was completed when PLF
visited the Mittag-Leffler Institute
in the Fall 2008 during the Semester Program ``Geometry, Analysis, and General Relativity''
organized by L. Andersson, P. Chrusciel, H. Ringstr\"om, and R. Schoen.  
P.A. was supported by the Portuguese Foundation for Science and Technology
(FCT) through grant SFRH/BD/17271/2004.  
The authors were supported by the A.N.R. (Agence Nationale de la Recherche)
through the grant 06-2-134423 entitled {\sl ``Mathematical Methods in General Relativity''} (MATH-GR), 
and by the Centre National de la Recherche Scientifique (CNRS).

%%%%%%%%%%%%%%%%%%%%%%%%%%%%%%%%%%%%%%%%%%%%%%%%%%%%%%%%%%%%%%%%%%%%%%%%%%%%%%%%%%%%%%%%%%%%%%%%%%%%%%%%%%%%%%

\end{document}